\documentclass[twocolumn]{autart}    % Enable this line and disable the
\usepackage{mathrsfs}
\usepackage{float}
\usepackage{amsfonts}
\usepackage{graphicx}          % Include this line if your
\usepackage{color}
\usepackage{amsmath}
\usepackage{caption}
\usepackage[titletoc,title]{appendix}
\usepackage[round]{natbib}

\newtheorem{remark}{Remark}

\newtheorem{assumption}{Assumption}
\newtheorem{lemma}{Lemma}

\newtheorem{proof of theorem 2}{Proof of Theorem 2}
\newtheorem{proof of theorem 1}{Proof of Theorem 1}

\begin{document}
\hyphenpenalty=5000
\tolerance=1000
\begin{frontmatter}

\title{Cooperative Extended State Observer Based Control of Vehicle Platoons With Arbitrarily Small Time Headway\thanksref{footnoteinfo}} % Title, preferably not more

\thanks[footnoteinfo]{
This paper was partially presented at the 21st IFAC World Congress, Berlin, Germany, July 12-17, 2020. Corresponding author: Tao Li. Tel. +86-21-54342646-318.
Fax +86-21-54342609.}
\author[Paestum]{Anquan Liu}\ead{anquanliu186@163.com},    % Add the
\author[Rome]{Tao Li}
%\thanks{* Corresponding author}
\ead{tli@math.ecnu.edu.cn}\textsuperscript{,*},               % e-mail address
\author[Rome]{Yu Gu}\ead{jessicagyrrr@126.com},  % (ead) as shown
\author[Rome]{Haohui Dai}\ead{hhdai@math.ecnu.edu.cn}

\address[Paestum]{School of Mechatronic Engineering and Automation, Shanghai University, Shanghai, 200072, China.}  % Please supply
\address[Rome]{Shanghai Key Laboratory of Pure Mathematics and Mathematical Practice,
	School of Mathematical Sciences, East China Normal University, Shanghai
	200241, China.}             % full addresses

                             % keyword list or with the% help of the Automatica
\begin{abstract}                          % Abstract of not more than 200 words.
We study platoon control of vehicles with linear third-order longitudinal dynamics under the constant time headway policy. The controller of each follower vehicle is only based on its own velocity, acceleration, inter-vehicle distance and velocity difference with respect to its immediate predecessor, which are all obtained by on-board sensors. We develop distributed cooperative extended state observers for followers to estimate the acceleration differences between adjacent vehicles. Based on estimates of the acceleration differences, distributed cooperative control laws are designed. By using the stability theory of perturbed linear systems, we show that the control parameters can be properly designed to ensure the closed-loop and $\mathcal{L}_2$ string stabilities for any given positive time headway. We further show that the proposed control law based on the ideal vehicle model can guarantee the closed-loop and $\mathcal{L}_2$ string stabilities even if there are small model parameter uncertainties. Also, simulation results demonstrate the robustness of the proposed control law against sensing noises, input delays and parameter uncertainties.
\end{abstract}

\end{frontmatter}

\section{Introduction}
  Vehicle platoon can improve road utilization rate and reduce fuel consumption effectively \citep{alam2011fuel}. Therefore, it has attracted worldwide attention \citep{shladover1991automated,coelingh2012all}. From networked control perspective, \cite{li2015overview} divided a vehicle platoon system into four basic modules: node dynamics, information flow topology, formation geometry, and distributed controller, among which the formation geometry greatly affects the stability of the vehicle platoon system. Formation geometry is determined by the spacing policy. The constant time headway policy is considered by  \cite{Klinge2009Time,Naus2010String,Xiao2011Practical,ploeg2014controller,Darbha2017Effects} and it is well known that a large time headway is conducive to string stabilities of vehicle platoon system \citep{Rajamani2002Semi,Naus2010String}, however, the larger the time headway is, the greater the inter-vehicle distance becomes, which leads to a lower road utilization rate.

It is of interest to guarantee the stability of the vehicle platoon with small time headways. Many  researches showed that using the accelerations of the preceding vehicles can reduce the lower bound of the time headway required \citep{Rajamani2002Semi,zhou2004string,Naus2010String,Darbha2017Effects,al2018feedforward}.
All the above works assumed that the accelerations of the preceding vehicles are obtained by the wireless communication network accurately, however, accurate communication doesn't exist in practical applications.
In addition, a control law which relies on communication data runs the risk of failure when the inter-vehicle communication network breaks down under attack.
Therefore, a cooperative control law with a small time headway, which can ensure both closed-loop  and string stabilities without relying on the inter-vehicle wireless communication network, is of especially significance for practical applications. \cite{Ploeg2015Graceful} and \cite{Wen2019Observer} proposed methods to estimate the acceleration differences between adjacent vehicles, respectively. The control laws in \cite{Ploeg2015Graceful} and \cite{Wen2019Observer} are indeed independent of wireless communication networks. The closed-loop and string stabilities are analyzed by numerical simulations in \cite{Ploeg2015Graceful} and \cite{Wen2019Observer}, especially,
\cite{Ploeg2015Graceful} considered a third-order linear model with input delays, in which the simulation results show that the string stability is not guaranteed if the time headway is small enough.

In this paper, we study platoon control of vehicles with linear third-order longitudinal dynamics under the constant time headway policy. We design a distributed cooperative control law for each follower vehicle only using its own velocity, acceleration, inter-vehicle distance and velocity difference with respect to its immediate predecessor, all of which can be obtained by on-board sensors. Firstly, the models of velocity difference between adjacent vehicles are established, based on which distributed cooperative extended state observers are designed to estimate the acceleration differences between adjacent vehicles. Then based on these estimates, distributed cooperative controllers are designed for follower vehicles. The controller for each follower consists of two parts, where the first part is a feedback term consisting of inter-vehicle distance error and its differential, the other is a feedforward term consisting of the estimate of the acceleration of the preceding vehicle. Thus, the whole cooperative control law only uses the data obtained by on-board sensors without wireless communication networks.

We analyze both closed-loop and $\mathcal{L}_2$ string stabilities of the vehicle platoon system. The closed-loop system matrix is decomposed into two matrices, one of which is related to feedback parameters of the distributed controllers, and the other is regarded as the perturbation matrix, which is related to feedforward parameters. Then, we give the range of the control parameters to ensure the closed-loop stability by using the stability theory of perturbed linear systems. In the frequency domain, we analyze the transfer functions which describe the inter-vehicle distance error propagation, and give the range of the control parameters to ensure the $\mathcal{L}_2$ string stability.  We show that one can design control parameters properly to ensure both closed-loop and $\mathcal{L}_2$ string stabilities for any given positive time headway. Our method for estimating the acceleration differences between adjacent vehicles is based on distributed cooperative extended state observers, which is totally different from those in \cite{Ploeg2015Graceful} and \cite{Wen2019Observer}. Besides, we give the explicit range of control parameters quantitatively related to the system parameters to ensure both closed-loop and $\mathcal{L}_2$ string stabilities for any given positive time headway.

We analyze the robustness of the proposed control law by both theoretical study and numerical simulations. (i) The proposed control law guarantees the exponential closed-loop stability, thus it is naturally robust against bounded sensing noises. (ii) We show that the proposed cooperative control law based on the ideal vehicle model can still ensure the closed-loop and $\mathcal{L}_2$ string stabilities provided the parameter uncertainties in the vehicle model are sufficiently small. (iii) In the numerical simulations, we consider the same vehicle model with input delays and the same time headway as \cite{Ploeg2015Graceful}, with additional sensing noises and uncertain model parameters. Simulation results show that the closed-loop and $\mathcal{L}_2$ string stabilities can be guaranteed by the proposed control law based on the ideal vehicle model.

The rest of this paper is organized as follows. The vehicle model and the control objectives are presented in Section II. In Section III, we first establish dynamic models based on velocity differences between every adjacent vehicles, then distributed cooperative extended state observers are designed to estimate the acceleration differences between adjacent vehicles. Finally, distributed cooperative controllers for follower vehicles are designed. In Section IV, we give the range of control parameters for the closed-loop and $\mathcal{L}_2$ string stabilities. In Section V, we analyze the robustness of the proposed control law against model parameter uncertainties. Numerical simulations are carried out in Section VI. In Section VII, we give some conclusions.

The following notation will be used throughout this paper. For a given matrix $A$, its 2-norm and minimum singular value are denoted by $\|A\|$ and $S_n(A)$, respectively; $diag(A)$ denotes a block diagonal matrix whose diagonal blocks are all matrix $A$; $\mathbb{C}$ denotes the complex domain; $\mathbb{R}$ denotes the real domain; $O$ and $I$ denote a zero matrix and an identity matrix with an appropriate size, respectively. For a given transfer function $G(s)$, its $\infty$-norm and inverse Laplace transform are  denoted by $\|G(s)\|_{\infty}$ and $\mathscr{L}^{-1}[G(s)]$, respectively.

\section{Problem formulation}
We consider the following third-order linear vehicle model which is commonly used in the vehicle platoon control \citep{Rajamani2002Semi,Ploeg2015Graceful,zheng2016stability,Wen2019Observer}.
\begin{align}\label{x_i}
\left\{
\begin{aligned}
&\dot p_i(t) = v_i(t), \vspace{1ex} \\
&\dot v_i(t) = a_i(t), \vspace{1ex} \hspace{20mm} i=0,1,...,N,\\
&\dot a_i(t) = -a_i(t)/\tau +u_i(t)/\tau,
\end{aligned}
\right.
\end{align}
where $p_i(t)$, $v_i(t)$ and $a_i(t)$ are the position, velocity and acceleration of the $i$th vehicle, respectively, $u_0(t)$ is the control input of the leader vehicle and $u_i(t)$ is the control input of the $i$th follower vehicle to be designed, $i=1,2,...,N$. The constant $\tau$ is the inertial delay of vehicle longitudinal dynamics. We consider the constant time headway spacing policy. The expected inter-vehicle distance is denoted by
\begin{align}\label{d_ri}
d_{r,i}(t)=r+h v_i(t),\;i=1,2,...,N,
\end{align}
where the constants $r$ and $h$ are the standstill distance and the time headway, respectively. The inter-vehicle distance error is given by,
\begin{align}\label{e_i}
e_i(t) = p_{i-1}(t)-p_i(t)-d_{r,i}(t),\;i=1,2,...,N,
\end{align}
which is the difference between the actual inter-vehicle distance and the expected inter-vehicle distance. Denote $G_{ei}(s)=\mathscr{E}_i(s)/\mathscr{E}_{i-1}(s)$, where $\mathscr{E}_i(s)$ is the Laplace transform of $e_i(t)$.
The control objectives are to design $u_i(t)$, $i=1,2,\ldots,N$, for follower vehicles so that the following two objectives are satisfied.

A. closed-loop stability: all the follower vehicles tend to move at the same velocity as the leader vehicle and the inter-vehicle distance errors converge to zero, i.e. $\lim \limits_ {t\to \infty}[v_i(t)-v_0(t)]=0$, $\lim \limits_ {t\to \infty}e_i(t)=0,i=1,2,\ldots,N$.

B. $\mathcal{L}_2 $ string stability: the inter-vehicle distance errors are not amplified during the backward propagation along the platoon, i.e. $\|G_{ei}(s)\|_{\infty}\leq 1,\;i=2,...,N.$

\section{Cooperative control law  based on extended state observer}
Suppose that there is no wireless communication between vehicles. We consider predecessor following topology, and each follower vehicle in the platoon relies on the on-board sensors to measure its own velocity, acceleration, the inter-vehicle distance and the velocity difference with respect to its immediate predecessor. However, the sensor of each follower vehicle cannot measure the acceleration of its immediate predecessor. Instead, we can design an observer to estimate the acceleration difference between adjacent vehicles. The extended state observer (ESO) was first put forward by \cite{han1995class}, whose core idea is to expand the unmodeled dynamics into new state and then according to the new state equation, an extended state observer is designed to estimate all states of the system. However, the extended state observer proposed by \cite{han1995class} is nonlinear, which has difficulty in parameter tuning and stability analysis. \cite{Gao2003Scaling} put forward a linear extended state observer, which simplifies parameter tuning and is also beneficial for stability analysis. In this paper, we design a linear extended state observer to estimate the acceleration difference between adjacent vehicles.

According to (\ref{x_i}), the models of the velocity difference between adjacent vehicles are given by
\begin{align}\label{v_di}
\left\{
\begin{aligned}
\dot v_{d,i}(t) &= a_{d,i}(t),  \vspace{1ex}\\
\dot a_{d,i}(t) &= q_i(t) +a_i(t)/\tau- u_i(t)/\tau, \vspace{1ex}\\
\dot q_i(t) &= w_i(t),\ i=1,2,...,N,
\end{aligned}
\right.
\end{align}
where
\vspace{-2mm}
\begin{align}
v_{d,i}(t) =& v_{i-1}(t)-v_i(t),\label{v_di1}  \vspace{1ex}\\
a_{d,i}(t) =& a_{i-1}(t)-a_i(t),\label{a_di1} \vspace{1ex} \\
q_i(t) =& (-a_{i-1}(t)+u_{i-1}(t))/\tau, \label{q_i} \vspace{1ex}\\
w_i(t) =& (a_{i-1}(t)-u_{i-1}(t)+\tau \dot u_{i-1}(t))/\tau^2.\label{w_i}
\end{align}
Here, $q_i(t)$ is the unmodeled dynamics, which contains the control input and the acceleration of $i-1$th vehicle that cannot be measured directly by the $i$th vehicle. We design a linear extended state observer
\begin{align}\label{z_i}
\left\{
\begin{aligned}
\dot z_{1,i}(t) = &z_{2,i}(t) +\beta_{1}(v_{d,i}(t)-z_{1,i}(t)), \vspace{1ex} \\
\dot z_{2,i}(t) = &z_{3,i}(t) +\beta_{2}(v_{d,i}(t)-z_{1,i}(t))+a_i(t)/\tau\\
&- u_i(t)/\tau, \vspace{1ex}  \\
\dot z_{3,i}(t) = &\beta_{3}(v_{d,i}(t)-z_{1,i}(t)),\ i = 1,2,\ldots,N, \\
\end{aligned}
\right.
\end{align}
%\begin{center}
%	\hspace{50mm}$i = 1,2,\ldots,N,$
%\end{center}
where $z_{2,i}(t)$ is the estimate of the acceleration difference $a_{d,i}(t)$ between the $i-1$th vehicle and the $i$th vehicle. The constants $\beta_{1}>0$, $\beta_{2}>0$ and $\beta_{3}>0$ are the observer gains to be designed. Then by the acceleration of $i$th vehicle, the estimate of the acceleration of $i-1$th vehicle is given by $z_{2,i}(t)+a_i(t)$. The controller of $i$th follower vehicle is designed as
\begin{align}\label{u_i1}
u_i(t) = &k_{p}e_i(t)+k_{v}(v_{d,i}(t)-ha_i(t)) \nonumber \\
&+k_{a}(z_{2,i}(t)+a_i(t)),\;i=1,2,...,N,
\end{align}
where $k_p>0$, $k_v>0$, $k_a>0$ are the control parameters to be designed. The controller (\ref{u_i1}) consists of two parts. The first part $k_{p}e_i(t)+k_{v}(v_{d,i}(t)-ha_i(t))$ is the feedback item, which consists of the inter-vehicle distance error between the adjacent vehicles and its differential; while the second part $k_{a}(z_{2,i}(t)+a_i(t))$ is the feedforward item, which consists of the estimate of the acceleration of the $i-1$th vehicle. It can be seen that the extended state observer (\ref{z_i}) and the controller (\ref{u_i1}) only use the information obtained by on-board sensors of followers.
\section{Stability analysis of vehicle platoon}
In reality, the velocity of the leader vehicle will reach at a steady state within a finite time $t_f$, that is, there exists $t_f$, such that $u_0(t)=0$, $t\geq t_f$. We make the following assumption.
\begin{assumption}
	$\lim \limits_ {t\to \infty }u_0(t)=0,\lim \limits_ {t\to \infty }\dot u_0(t)=0$.
\end{assumption}
\vspace{-3mm}		
Denote $\Psi(k_p,k_v) = {
	\left[ \begin{array}{cc}
	\Psi_{11} & O \\
	\Psi_{21} & \Psi_{22}
	\end{array}
	\right ]}\in \mathbb{R}^{6N\times6N}$,
where $\Psi_{11}$, $\Psi_{21}$ and $\Psi_{22}$ are $3N$ dimensional matrices with $\Psi_{22} = diag(\mathcal{H})$,
\begin{equation}
\mathcal{A}
=\begin{bmatrix}
%\begin{smallmatrix}
0 & 1 & -h\\
0 & 0 & -1\\
\begin{aligned}k_p/\tau\end{aligned} & \begin{aligned}k_v/\tau\end{aligned} &\begin{aligned}(-1-k_vh)/\tau\end{aligned}
%\end{smallmatrix}
\end{bmatrix} \nonumber,
\mathcal{B}
= \begin{bmatrix}
0 & 0 & 0\\
0 & 0 & 1 \\
0 & 0 & 0
\end{bmatrix}, \nonumber
\end{equation}
\begin{equation}
\mathcal{E}
=\begin{bmatrix}
0 & 0 & 0\\
0 & 0 & 0 \\
0 & 0 & \begin{aligned}-k_v/\tau\end{aligned}
\end{bmatrix} \nonumber,
\mathcal{H}
= \begin{bmatrix}
-\beta_1  &  1  & 0 \\
-\beta_2  &  0  & 1 \\
-\beta_3  &  0  & 0    %\substack{i=1 \\ j=1 \\j = 1}
\end{bmatrix}, \nonumber
\end{equation}
\begin{equation}
\mathcal{D}
= \begin{bmatrix}
0 & 0 & 0\\
0 & 0 & 0\\
\begin{aligned}&k_p(1+k_vh)\\&/\tau^2\end{aligned} & \begin{aligned}&(k_v+k_v^2h\\&-k_p\tau)/\tau^2\end{aligned} & \begin{aligned}&(k_ph\tau+k_v\tau-2k_vh\\&-k_v^2h^2-1)/\tau^2\end{aligned}
\end{bmatrix}.  \nonumber
\end{equation}
\begin{equation}
\Psi_{11}
=\begin{bmatrix}
\mathcal{A} & \quad & \quad & \quad  \\
\mathcal{B} & \mathcal{A} & \quad & \quad \\
\quad & \ddots & \ddots & \quad  \\
\quad & \quad & \mathcal{B} & \mathcal{A}
\end{bmatrix}, \nonumber
\Psi_{21}
=\begin{bmatrix}
O & \quad & \quad & \quad & \quad  \\
\mathcal{D} & O & \quad & \quad & \quad \\
\mathcal{E} & \mathcal{D} & O & \quad & \quad \\
\quad & \ddots & \ddots & \ddots & \quad \\
\quad  & \quad & \mathcal{E} & \mathcal{D} & O
\end{bmatrix},\nonumber
\end{equation}
For the closed-loop and $\mathcal{L}_2$ string stabilities, we have the following theorems.
\begin{thm}\label{theorem1}
	Suppose that Assumption 1 holds. Consider the system (\ref{x_i}) under the control law (\ref{z_i}) and (\ref{u_i1}). For any given $h>0$, if $\beta_1>0$, $\beta_3>0$, $\beta_1\beta_2-\beta_3>0$, $k_p>0$, 	
	\begin{equation}\label{h3}
	k_v>
	\left\{
	\begin{array}{ll}
	\begin{aligned}&(((1-k_ph^2)^2+4k_ph\tau)^{1/2}-(1+k_ph^2))\\
                   &/(2h),\;\hspace{42.5mm} {\rm if} \ h<\tau,\vspace{1ex}  \end{aligned}\\
	\begin{aligned}    0, \; \hspace{49mm}{\rm if} \ h\geq\tau,\end{aligned}
	\end{array}
	\right.
	\end{equation}
	\begin{equation}\label{k_a}
	0<k_a<
	\left\{
	\begin{array}{ll}
	\begin{aligned}r_c(\Psi)\tau \end{aligned},\; \hspace{38.5mm}{\rm if} \ N=1,\vspace{1ex}\\
	\begin{aligned}&r_c(\Psi) \tau^2/(k_vh+\tau\beta_2+4\tau+1),\end{aligned}\;{\rm if} \ N=2, \vspace{1ex}\\
	\begin{aligned}&(\tau(\Theta^2\tau^2+4(2N-5)r_c(\Psi))^{1/2}-\tau^2\Theta)\\
                   &/(2(2N-5)),\; \hspace{29mm} {\rm if}  \  N\geq3,\end{aligned}
	\end{array}
	\right.
	\end{equation}
    where $\Theta = (N\tau+(N-1)(\tau\beta_2+2\tau+k_vh+1)+(N-2)(k_p+k_v+2k_vh+2))/\tau^2$, $r_c(\Psi) = \min \limits_{\omega \in \mathbb{R}}S_n(i\omega I- \Psi)$,then $v_i(t)-v_0(t)$ and $e_i(t)$ both converge to zero exponentially, $i=1,2,...,N$.
\end{thm}
\vspace{-2mm}
The proof of Theorem \ref{theorem1} is given in Appendix \ref{Theorem 1}.
\vspace{-2mm}
\begin{remark}\label{Remark1}
   \emph{Theorem \ref{theorem1} shows that the control law (\ref{z_i}) and (\ref{u_i1}) can be properly designed such that the closed-loop system is exponentially stable, so the proposed control law is naturally robust against sensing noises, i.e. under the control law (\ref{z_i}) and (\ref{u_i1}) with bounded sensing noises in the measurements of $v_{d,i}(t)$,  $e_i(t)$ and $a_i(t)$, the inter-vehicle distance errors will converge to a neighborhood of zero whose size is proportional to the noise intensities. }
\end{remark}

%For the $\mathcal{L}_2$ string stability of the platoon, we have the following theorem.
\vspace{-2mm}
\begin{thm}\label{theorem2}
	Consider the system (\ref{x_i}) under the control law (\ref{z_i}) and (\ref{u_i1}). Let $k_p = \mu_pk$, $k_v=\mu_vk$, $k_a=\mu_ak$, $\beta_1=3\omega_o$, $\beta_2=3\omega_o^2$, $\beta_3=\omega_o^3$, where $k$, $\mu_p$, $\mu_v$, $\mu_a$, $\omega_o$ are positive parameters to be designed. For any given $h>0$, if $\mu_a>0$, $\mu_p> 0$,
		\begin{align}
		\mu_v>&\max\left\{\sqrt{3}\mu_a/h,2\mu_a/h^2\right\}\label{mu_v},\\
		\omega_o > &16\mu_v\mu_a/(3h^2\mu_v^2-9\mu_a^2),\label{omega}\\
		k \geq&\max\left\{ \theta_1, \theta_2, \theta_3, \theta_4,\gamma_5/\alpha_5 \right\}\label{k},
		\end{align}
where the definitions of $\theta_i,i=1,2,3,4$, $\alpha_5$ and $\gamma_5$ are given in Appendix \ref{parameters}, then $\|G_{ei}(s)\|_{\infty}\leq 1,\ i=2,...,N$.
\end{thm}	
\vspace{-3mm}	
The proof of Theorem \ref{theorem2} is given in Appendix \ref{Theorem 2}.

\begin{remark}\label{Remark6}
\emph{The control parameters can be properly designed to guarantee closed-loop and $\mathcal{L}_2$ string stabilities simultaneously. From Theorem \ref{theorem2}, we know $\lim _{\mu_v\to +\infty\atop\mu_a\to 0}\max\{\theta_1,\theta_2,\theta_3,\theta_4,\gamma_5/\alpha_5\}=2/(h^2\mu_p)$. We first fix $\mu_p>0$ and $k>2/(h^2\mu_p)$. It is known that there are $\mu_v^*$ and $\mu_a^*$ such that (\ref{k}) holds for any $\mu_v>\mu_v^*$ and $\mu_a<\mu_a^*$. Then choose $\mu_v\in(\mu_v^*,+\infty)$ and $\mu_a\in(0,\mu_a^*)$ such that (\ref{h3}) and (\ref{k_a}) hold, respectively.
}
\end{remark}
\begin{remark}\label{Remark2}
\emph{Theorem \ref{theorem2} shows that the control parameters can be properly designed to ensure the $\mathcal{L}_2$ string stability for any given positive time headway. There is another definition of string stability, called $\mathcal{L}_{\infty}$ string stability, which means $\sup_{t\geq0}|e_i(t)|\leq\sup_{t\geq0}|e_{i-1}(t)|,i=2,...,N$.
According to \cite{J1998A}, if $\|G_{ei}(s)\|_{\infty}\leq1$ and $\mathscr{L}^{-1}[G_{ei}(s)]\geq0$, then $\mathcal{L}_{\infty}$ string stability is satisfied.}
\end{remark}
\begin{remark}\label{Remark4}
\emph{In reality, the velocity of leader vehicle is positive. According to \cite{J1998A}, if $\|G_{vi}(s)\|_{\infty}\leq 1$ and $\mathscr{L}^{-1}[G_{vi}(s)]\geq0$, where $G_{vi}(s)=\mathscr{V}_i(s)/\mathscr{V}_{i-1}(s)$, $\mathscr{V}_i(s)$ is Laplace transform of $v_i(t)$, then $\sup_{t\geq0}|v_i(t)|\leq\sup_{t\geq0}|v_{i-1}(t)|,i=1,...,N$, and the $v_i(t),i=1,...,N$ have the same sign as $v_0(t)$. Thus
the velocities of follower vehicles are positive. The evolution of $|G_{vi}(j\omega)|$ and $\mathscr{L}^{-1}[G_{vi}(s)]$ are investigated by numerical simulation in Section 6.
}
\end{remark}

\section{Robustness against parameter uncertainties}
\vspace{-2mm}
 The control law (\ref{z_i}) and (\ref{u_i1}) is designed based on the ideal vehicle model (\ref{x_i}) with completely known parameters. In practice, the vehicle model may be inaccurate. In this section, we  will give the range of control parameters to ensure the closed-loop and $\mathcal{L}_2$ string stabilities when there are parameter uncertainties in the vehicle model. It is shown that the control law (\ref{z_i}) and (\ref{u_i1}) based on the ideal vehicle model can ensure the closed-loop and $\mathcal{L}_2$ string stabilities with small model parameter uncertainties.

Suppose that the real vehicles have the following dynamics
\begin{align}\label{x_ie}
\left\{
\begin{aligned}
&\dot p_i(t) = v_i(t), \vspace{1ex} \\
&\dot v_i(t) = a_i(t), \vspace{1ex} \hspace{25mm} i=0,1,...,N,\\
&\dot a_i(t) = -\left(1/\tau+\epsilon_i\right) a_i(t)+\left(1/\tau+\epsilon_i\right)u_i(t),
\end{aligned}
\right.
\end{align}
where the definitions of $u_i(t),i=0,1,...,N$ are the same as in (\ref{x_i}). The constant $\tau$ is the known nominal inertial delay of vehicle longitudinal dynamics. The constant $\epsilon_i$ is the parameter uncertainty. We assume that $|\epsilon_i|\leq \overline \epsilon$, $i=0,1,...,N$, where $ \overline\epsilon\in[0,1/\tau)$. We have the following theorems.
\vspace{-2mm}
\begin{thm}\label{theorem3}
	Suppose that Assumption 1 holds. Consider the system (\ref{x_ie}) under the control law (\ref{z_i}) and (\ref{u_i1}). For any given $h>0$, if $\beta_1>0$, $\beta_3>0$, $\beta_1\beta_2-\beta_3>0$, $k_p>0$, 	
	\begin{align}\label{h3e}
	k_v>
	\left\{
	\begin{array}{ll}
	\begin{aligned}&(((1-k_ph^2)^2+4k_ph\tau)^{1/2}-(1+k_ph^2))\\
                   &/(2h),\; \hspace{42.5mm}{\rm if} \ h<\tau,\vspace{1ex}  \end{aligned}\\
	\begin{aligned}    0, \; \hspace{49mm}{\rm if} \ h\geq\tau,\end{aligned}
	\end{array}
	\right.
	\end{align}
    \vspace{-8mm}
\begin{align}\label{oe1}
        \left\{
        \begin{array}{lc}
        \begin{aligned}&Z_1(k_p,k_v)\overline\epsilon^2+Z_2(k_p,k_v)\overline\epsilon-r_c(\Psi(k_p,k_v))<0,\\\;&\hspace{65mm}{\rm if} \ N=1,\end{aligned} \vspace{1ex}\\
        \begin{aligned}&Z_3(k_p,k_v)\overline \epsilon^2+Z_4(k_p,k_v)\overline \epsilon-r_c(\Psi(k_p,k_v))<0,\\\;&\hspace{65mm}{\rm if} \ N=2,\end{aligned} \vspace{1ex}\\
        \begin{aligned}
        &Z_5(k_p,k_v)\overline \epsilon^2+Z_6(k_p,k_v)\overline \epsilon-r_c(\Psi(k_p,k_v))<0,\\\;&\hspace{65mm}{\rm if} \ N\geq3,
        \end{aligned}
        \end{array}
        \right.
    \end{align}
     \vspace{-8mm}
    \begin{align}\label{k_ae}
    0<k_a <\left\{
    \begin{array}{ll}
    \begin{aligned}&\left(r_c(\Psi)-Z_1\overline\epsilon^2-Z_2\overline\epsilon\right)/(Y_1\overline\epsilon^2+Y_2\overline\epsilon+1/\tau)\\&\hspace{51mm}{\rm if} \ N=1, \end{aligned} \vspace{1ex}\\
     \begin{aligned}
    &[((\Theta_1+Y_3\overline \epsilon^2+Y_4\overline \epsilon)^2+4A_1(r_c(\Psi)-Z_3\overline\epsilon^2\\
    &-Z_4\overline\epsilon))^{1/2}-(\Theta_1+Y_3\overline \epsilon^2+Y_4\overline \epsilon)]/(2A_1),\\&\hspace{51mm}{\rm if} \ N=2,
    \end{aligned} \vspace{1ex}\\
    \begin{aligned}
    &[((\Theta+Y_5\overline \epsilon^2+Y_6\overline \epsilon)^2+4A_2(r_c(\Psi)-Z_5\overline\epsilon^2\\
    &-Z_6\overline\epsilon))^{1/2}-(\Theta+Y_5\overline \epsilon^2+Y_6\overline \epsilon)]/(2A_2),\\&\hspace{51mm}{\rm if} \ N\geq3,
    \end{aligned}
    \end{array}
    \right.
	\end{align}
     where the definitions of $Z_i,Y_i,i=1,2,...,6$, $\Theta_1$, $A_1$ and $A_2$ are given in Appendix \ref{parameters}, then $v_i(t)-v_0(t)$ and $e_i(t)$ both converge to zero exponentially, $i=1,2,...,N$.
\end{thm}
\vspace{-2mm}
The proof of Theorem \ref{theorem3} is given in Appendix \ref{Theorem 3}.
\vspace{-2mm}
\begin{thm}\label{theorem4}
	Consider the system (\ref{x_ie}) under the control law (\ref{z_i}) and (\ref{u_i1}). Let $k_p = \mu_pk$, $k_v=\mu_vk$, $k_a=\mu_ak$, $\beta_1=3\omega_o$, $\beta_2=3\omega_o^2$, $\beta_3=\omega_o^3$, where $k$, $\mu_p$, $\mu_v$, $\mu_a$, $\omega_o$ are positive parameters to be designed. For any given $h>0$, if $\mu_a>0$, $\mu_p> 4\mu_a\overline b/(\tau\underline b^2h^2)$,
		\begin{align}
		\mu_v>&\max\left\{4\mu_a\overline b^2/(h\underline b^2),2\mu_a\overline b/(\tau h\underline b^2)\right\}\label{mu_ve},\\
		\omega_o > &\max\left\{(|\underline \lambda_{2}^2-4\underline \lambda_{1}\underline \lambda_{3}|^{1/2}-\underline \lambda_{2})/(2\underline \lambda_{1}),(3(2\underline b\mu_p\mu_a/\tau\right.\nonumber\\
&\hspace{8mm}\left.+2\underline b^2\mu_p\mu_a+\underline b^2h^2\mu_p^2)/(\underline b^2h^2\mu^2_v-\overline b^2\mu^2_a))^{1/2}
	 \right\},\label{omegae}\\
		k \geq&\max\left\{ \theta_{1}, \theta_{2}, \theta_{3}, \theta_{4},\overline \gamma_5/\underline \alpha_5 \right\}\label{ke},
		\end{align}
where the definitions of $\overline b$, $\underline b$, $\underline \alpha_5$, $\overline \gamma_5$, $\theta_i,i=1,2,3,4$ and $\underline \lambda_i,i=1,2,3$ are given in Appendix \ref{parameters}, then $\|G_{ei}(s)\|_{\infty}\leq 1, \ i=2,...,N$.
\end{thm}
\vspace{-4mm}		
The proof of Theorem \ref{theorem4} is given in Appendix \ref{Theorem 4}.
\vspace{-3mm}
\begin{remark}
\emph{From Theorem \ref{theorem3}, we know that the control law (\ref{z_i}) and (\ref{u_i1}) is robust against model parameter uncertainties. If $\overline \epsilon=0$, then
(\ref{oe1}) holds and (\ref{k_ae}) degenerates into (\ref{k_a}). By the continuity of (\ref{oe1}) and (\ref{k_ae}) with respect to  $\overline \epsilon$, we know that if the control law (\ref{z_i}) and (\ref{u_i1}) is designed according to Theorem 1 based on the ideal vehicle model (\ref{x_i}), then the closed-loop and $\mathcal{L}_2$ string stabilities can be ensured provided $\overline \epsilon$ is sufficiently small. }
\end{remark}
\vspace{-3mm}
\begin{remark}\label{Remark5}
\emph{A more accurate nonlinear vehicle model is considered in \cite{zheng2016stability}
\begin{align}\label{T_i}
\left\{
\begin{aligned}
\dot p_i(t) = &v_i(t),  \vspace{1ex}\\
\dot v_i(t) = &a_i(t),  \hspace{28mm} i = 0,1,\ldots,N,\vspace{1ex} \\
\dot a_i(t) = &\eta_iT_{i,des}(t)/(m_iR_i\tau_i)-(2C_iv_i(t)a_i(t)\tau_i\vspace{1ex}\\
              &+m_ia_i(t)+C_iv_i^2(t)+m_igf_i)/(m_i\tau_i),
\end{aligned}
\right.
\end{align}
where $T_{i,des}(t)$ is the expected driving or braking torque of the $i$th vehicle at time $t$. The constant $m_i$, $f_i$, $R_i$, $C_i$, $\tau_i$ and $\eta_i$ are the mass, the rolling resistance coefficient, the tire radius, the total air resistance coefficient, the inertial delay of longitudinal dynamics and the mechanical efficiency of the drive train of the $i$th vehicle, respectively. The constant $g$ is the gravitational acceleration. The feedback linearization law is given by
\begin{align}\label{T_i,des}
T_{i,des}(t)= &\widetilde R_i[\widetilde C_i(v_i(t)+\xi_i^v(t))(2\widetilde \tau_i (a_i(t)+\xi_i^a(t)) \nonumber  \\
&+(v_i(t)+\xi_i^v(t)))+\widetilde m_ig\widetilde f_i+\widetilde m_iu_i(t)]/\widetilde \eta_i,\nonumber \\
&\hspace{30mm}\;i = 0,1,\ldots,N.
\end{align}
where the constants $\widetilde m_i$, $\widetilde f_i$, $\widetilde R_i$, $\widetilde C_i$, $\widetilde \eta_i$ and $\widetilde \tau_i$ are the nominal parameters of $i$th vehicle, $v_i(t)+\xi_i^v(t)$ and $a_i(t)+\xi_i^a(t)$ are the velocity and acceleration of $i$th vehicle measured by on-board sensors, $\xi_i^v(t)$ and $\xi_i^a(t)$ are sensing noises . Substituting (\ref{T_i,des}) into (\ref{T_i}), we obtain
\begin{align}\label{model2}
\left\{
\begin{aligned}
\dot p_i(t) = &v_i(t),  \vspace{1ex}\\
\dot v_i(t) = &a_i(t),  \hspace{22mm} i = 0,1,\ldots,N,\\
\dot a_i(t) = &\zeta_i(t)- (a_i(t)+\xi_i^a(t))/\widetilde \tau_i+u_i(t)/\widetilde \tau_i,
\end{aligned}
\right.
\end{align}
where $\zeta_i(t)=\eta_iT_{i,des}(t)/(m_iR_i\tau_i)-(2C_iv_i(t)a_i(t)\tau_i+C_iv_i^2(t)+m_igf_i)/(m_i\tau_i)+ (a_i(t)+\xi_i^a(t))/\widetilde \tau_i- u_i(t)/\widetilde \tau_i$. Compared with (\ref{x_i}), the model (\ref{model2}) contains a nonlinear term $\zeta_i(t)$. From (\ref{model2}), we know the models of the velocity difference between adjacent vehicles (\ref{v_di}) with $q_i(t) = \zeta_{i-1}(t)-\zeta_i(t)+(-( a_{i-1}(t)+\xi_{i-1}^a(t))+u_{i-1}(t))/\widetilde\tau_i$. The ESO (\ref{z_i}) can still be used to estimate the acceleration differences between adjacent vehicles. The nonlinear term $\zeta_i(t)$ can be estimated by designing another ESO
\begin{align}\label{z_i1}
\left\{
\begin{aligned}
\dot {\hat v}_i(t) = &{\hat a}_i(t) +\beta_{4}(v_i(t)+\xi_i^v(t)-\hat v_i(t)), \vspace{1ex} \\
\dot {\hat a}_i(t) = &{\hat \zeta}_i(t) +\beta_{5}(v_i(t)+\xi_i^v(t)-\hat v_i(t))\\
&-(a_i(t)+\xi_i^a(t))/\widetilde\tau_i+u_i(t)/\widetilde\tau_i, \vspace{1ex}  \\
\dot {\hat \zeta}_i(t) = &\beta_{6}(v_i(t)+\xi_i^v(t)-\hat v_i(t)),\ i = 1,\ldots,N, \\
\end{aligned}
\right.
\end{align}
where $\hat v_i(t)$, $\hat a_i(t)$ and  $\hat \zeta_i(t)$ are the estimates of $v_i(t)$, $a_i(t)$ and $\zeta_i(t)$, respectively. The constants $\beta_4>0$, $\beta_5>0$ and $\beta_6>0$ are the observer gains to be designed.
The controller of $i$th follower vehicle is designed as
\begin{align}\label{u_xi}
u_i(t) = &k_{p}(e_i(t)+\xi_i^e(t))+k_{v}(v_{d,i}(t)+\xi_i^d(t)-h(a_i(t)\nonumber\\
&+\xi_i^a(t)))+k_a(z_{2,i}(t)+a_i(t)+\xi_i^a(t))\nonumber\\
&-\widetilde \tau_i\hat \zeta_i(t),\;i=1,...,N,
\end{align}
where $e_i(t)+\xi_i^e(t)$ and $v_{d,i}(t)+\xi_i^d(t)$ are the inter-vehicle distance error and velocity difference between adjacent vehicles measured by on-board sensors, $\xi_i^e(t)$ and $\xi_i^d(t)$ are sensing noises.
The stability analysis of the vehicle platoon system (\ref{model2}) under control law (\ref{z_i}), (\ref{z_i1}) and (\ref{u_xi}) would be an interesting and challenging topic for future investigation.
}\end{remark}

\vspace{-2mm}
\section{NUMERICAL SIMULATIONS}
Suppose there are 1 leader vehicle and 5 follower vehicles with the following dynamics with parameter uncertainties
\begin{align*}
\left\{
\begin{aligned}
&\dot p_i(t) = v_i(t), \vspace{1ex} \\
&\dot v_i(t) = a_i(t), \vspace{1ex} \hspace{35mm} i=0,1,...5,\\
&\dot a_i(t) = -\left(1/\tau+\epsilon_i\right) a_i(t)+\left(1/\tau+\epsilon_i\right)u_i(t),
\end{aligned}
\right.
\end{align*}
where $\tau=0.1$. The parameter uncertainties are given by $\epsilon_0=-0.8$, $\epsilon_1=0.1$, $\epsilon_2=0.5$, $\epsilon_3=-0.2$, $\epsilon_4=0.65$, $\epsilon_5=-0.3$.

\begin{figure}[H]
	\centering
	\begin{minipage}[t]{0.236\textwidth}
		\centering
		\includegraphics[width=4.45cm]{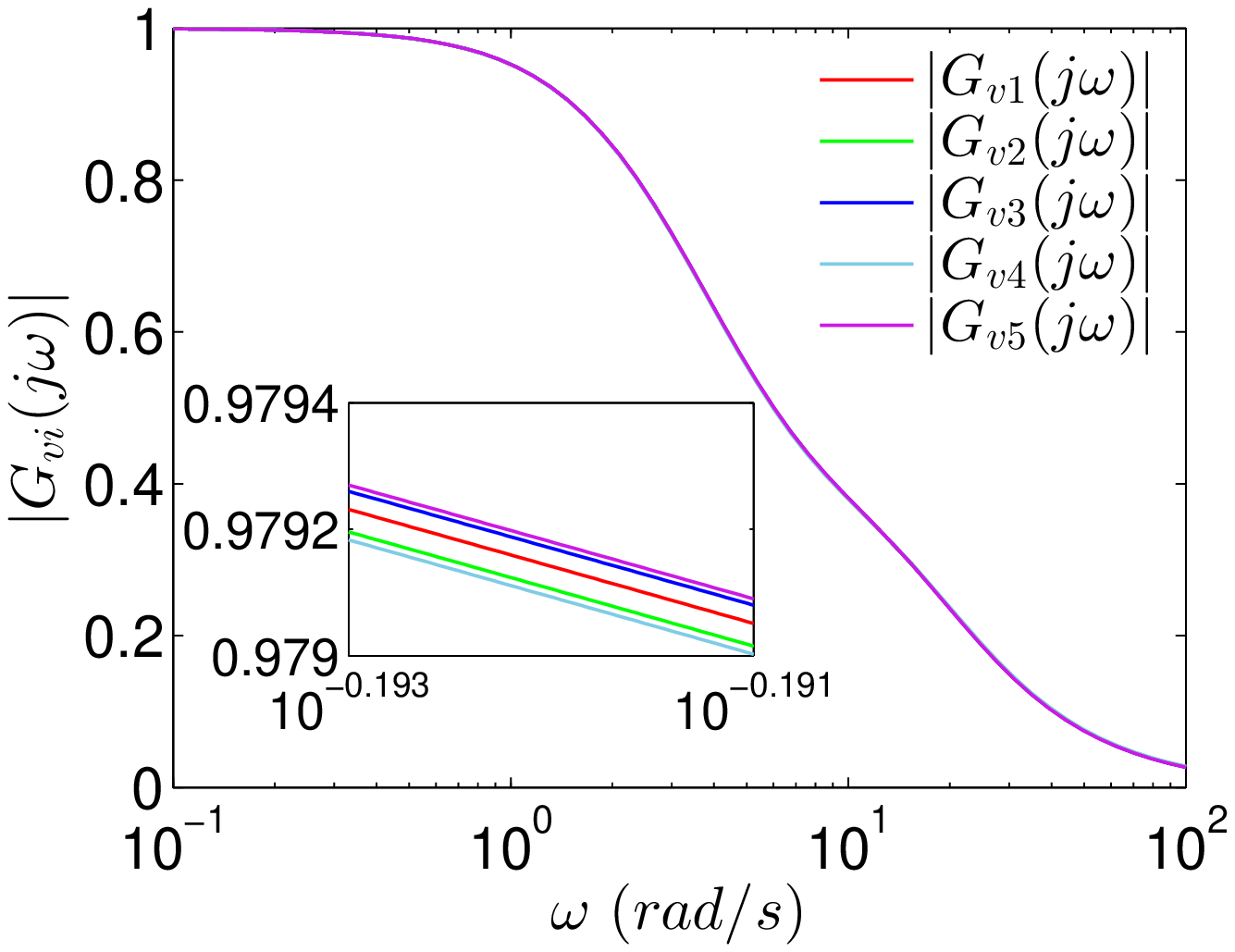}
		\caption{(a)}
\label{FIG1}
	\end{minipage}
	\begin{minipage}[t]{0.236\textwidth}
		\centering
		\includegraphics[width=4.45cm]{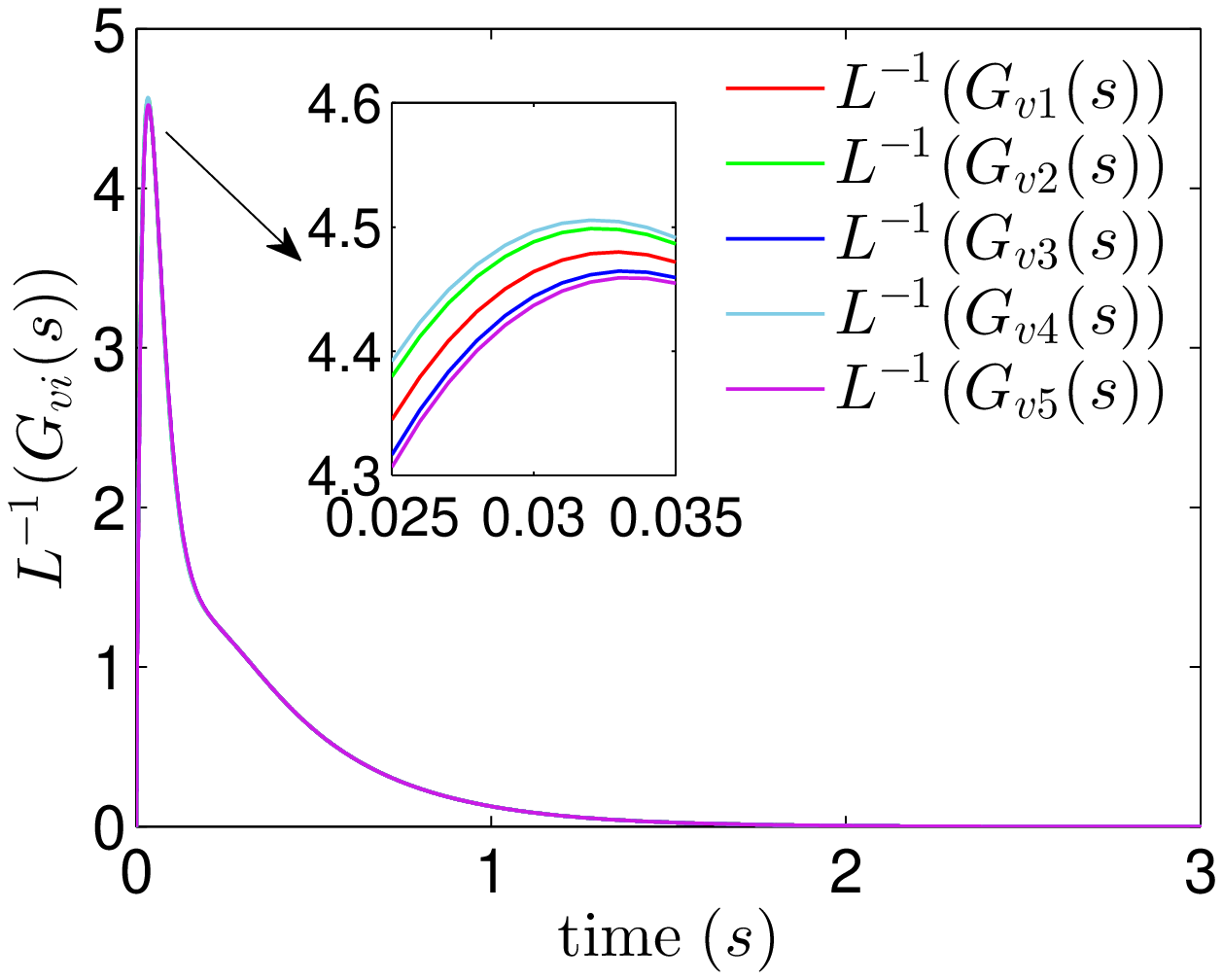}
		\caption*{(b)}
	\end{minipage}
\end{figure}

To make the behavior of the leader vehicle close to a real vehicle, we construct a virtual vehicle in CarSim and record its acceleration. Then we use the recorded acceleration as the expected acceleration of the leader vehicle in the following simulations. The initial velocities are given by $v_i(0)=10\,m/s$, $i = 0,1,...,5$. The initial accelerations are given by $a_i(0)=0\,m/s^2$, $i = 0,1,...,5$. The initial positions are taken as $p_0(0)=30\,m$, $p_1(0)=24\,m$, $p_2(0)=18\,m$, $p_3(0)=12\,m$, $p_4(0)=6\,m$, $p_5(0)=0\,m$. The standstill distance is given by $r=3\,m$. Let $h=0.3s$. We choose $k_p=8$, $k_v=40$, $k_a=1.2$ and $\omega_o=15$, $\beta_1=45$, $\beta_2 = 675$ and $\beta_3=3375$. The evolution of $|G_{vi}(j\omega)|$ and $\mathscr{L}^{-1}[G_{vi}(s)]$ are shown in Fig.\ref{FIG1}.

\begin{figure}[H]
	\centering
	\begin{minipage}[t]{0.236\textwidth}
		\centering
		\includegraphics[width=4.5cm]{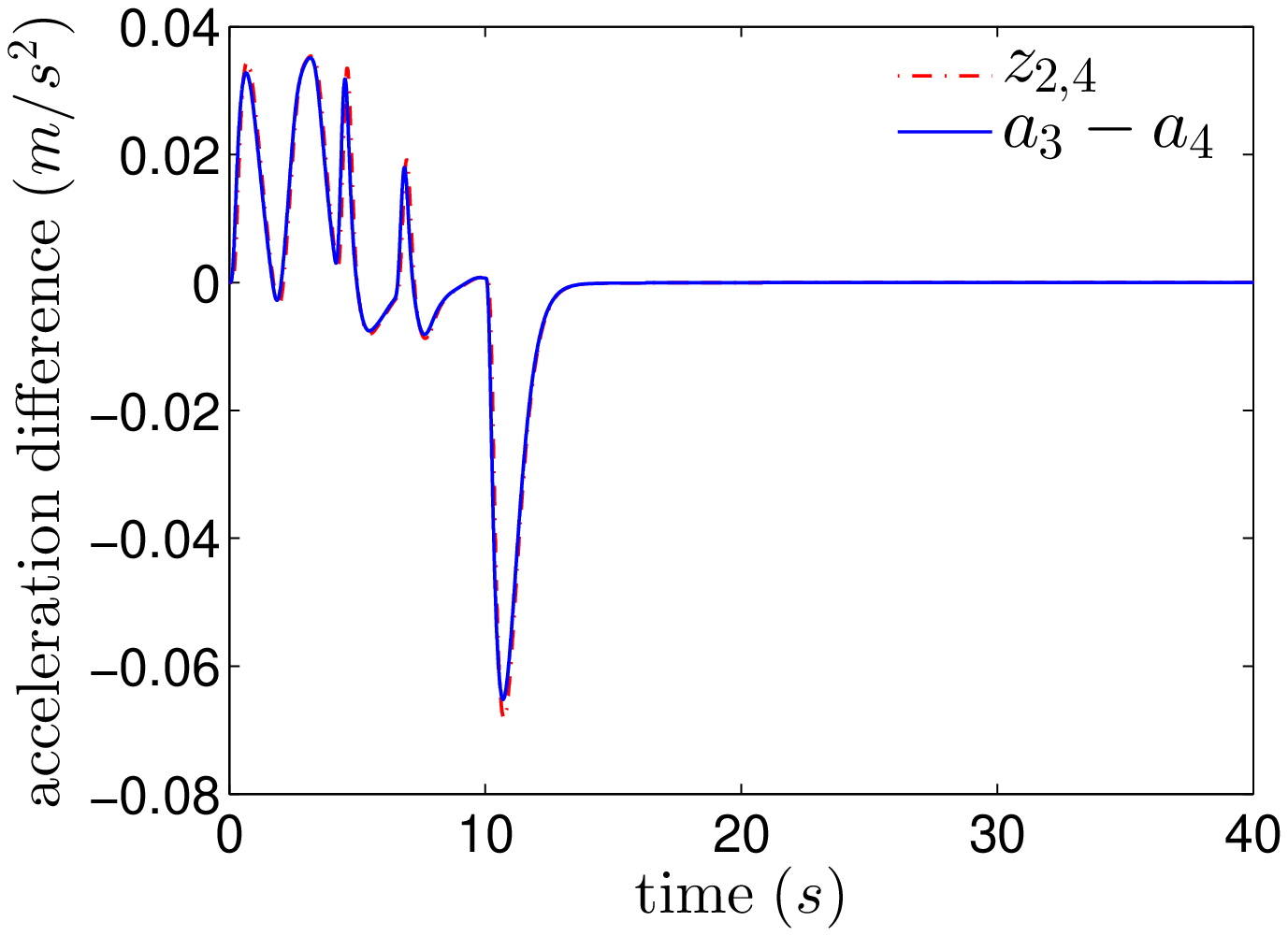}
		\caption{(a)}
\label{FIG2}
	\end{minipage}
	\begin{minipage}[t]{0.236\textwidth}
		\centering
		\includegraphics[width=4.5cm]{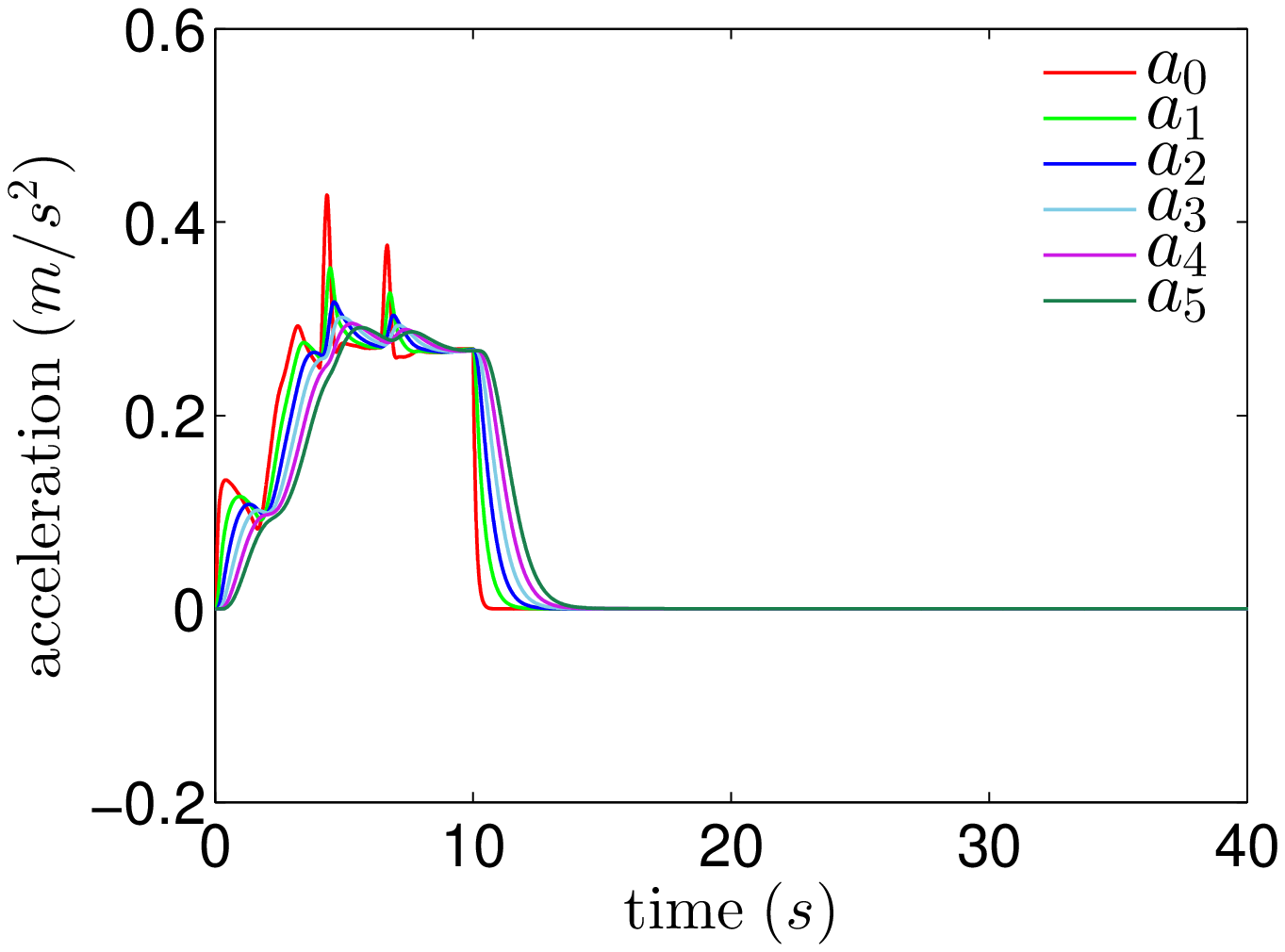}
		\caption*{(b)}
	\end{minipage}
	\centering
	\begin{minipage}[t]{0.236\textwidth}
		\centering
		\includegraphics[width=4.5cm]{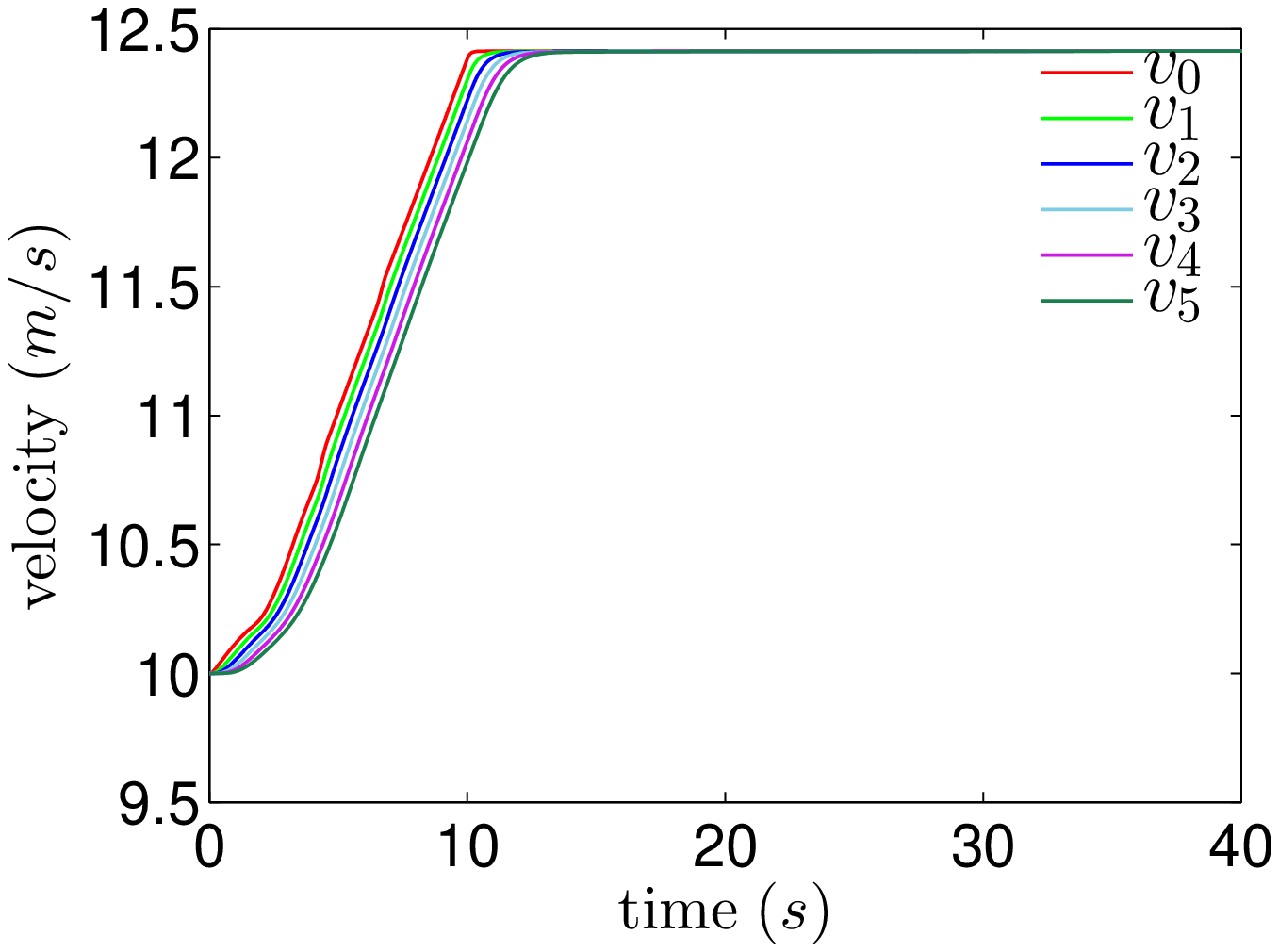}
		\caption*{(c)}
	\end{minipage}
	\begin{minipage}[t]{0.236\textwidth}
		\centering
		\includegraphics[width=4.5cm]{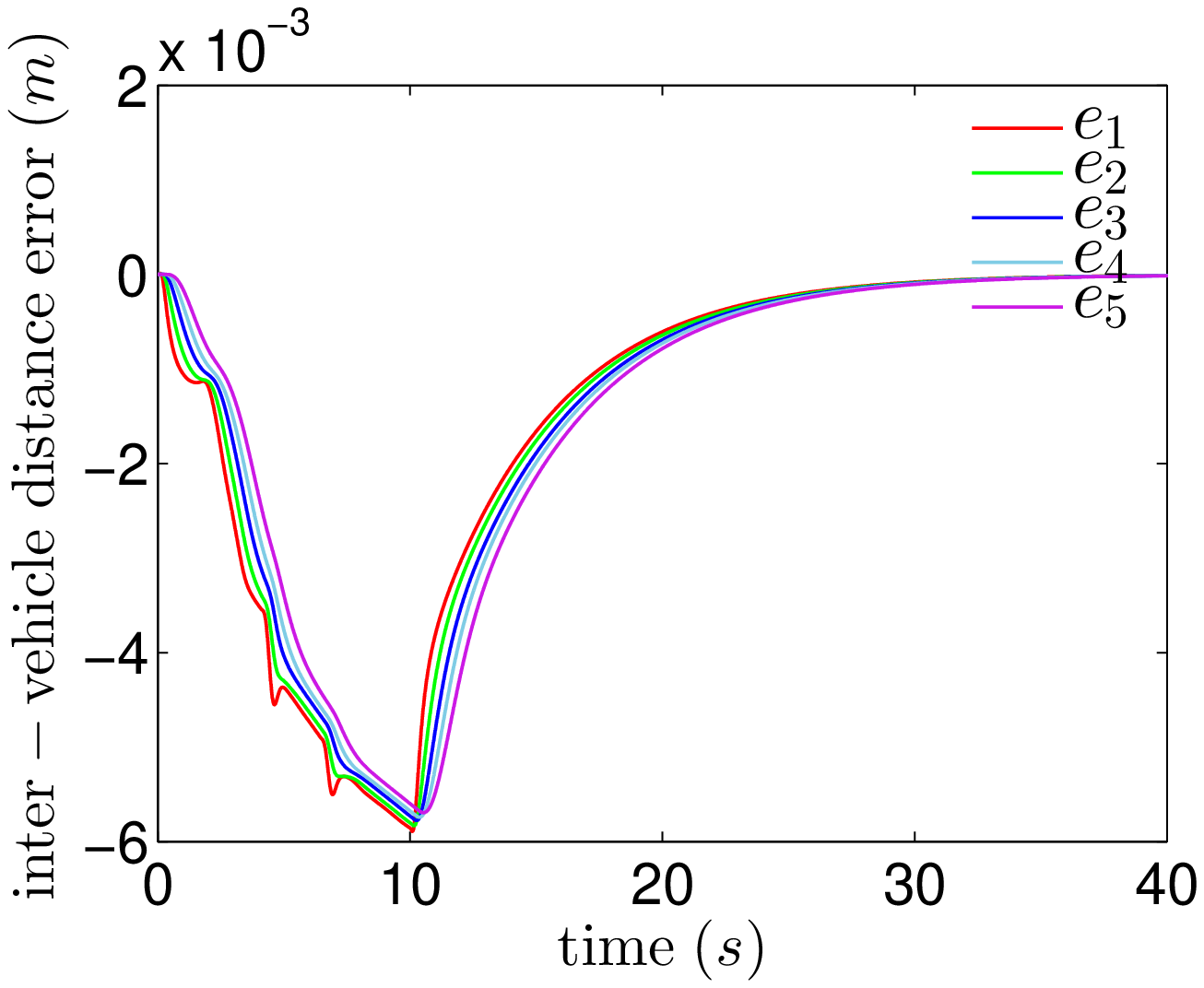}
		\caption*{(d)}
	\end{minipage}
	\caption* {Vehicle model with parameter uncertainties. (a) The actual and the estimated acceleration differences between the 3rd and 4th follower vehicles. (b) Accelerations of the vehicles. (c)Velocities of the vehicles.  (d) Inter-vehicle distance errors.}
\end{figure}

The actual and the estimated acceleration differences between the 3rd and 4th follower vehicles are shown in Fig.\ref{FIG2}(a). The evolution of vehicles' accelerations and velocities are shown in Fig.\ref{FIG2}(b) and Fig.\ref{FIG2}(c), respectively. The evolution of inter-vehicle distance errors are shown in Fig.\ref{FIG2}(d).

Fig.\ref{FIG1} shows that $\|G_{vi}(s)\|_{\infty}\leq 1$ and $\mathscr{L}^{-1}[G_{vi}(s)]\geq0,i=1,2,...,5$. Then from Remark \ref{Remark4}, we know that $\sup_{t\geq0}|v_i(t)|\leq\sup_{t\geq0}|v_{i-1}(t)|$ and the $v_i(t),i=1,...,5$ have the same sign as $v_0(t)$, which are also shown in Fig.\ref{FIG2}.(c). From Fig. \ref{FIG2}.(a), it is shown that although the ESO (\ref{z_i}) are designed based on the nominal $\tau$, the acceleration differences between adjacent vehicles $a_{d,i}(t)$ can be estimated well by the ESO (\ref{z_i}). Fig.\ref{FIG2}.(b) and Fig.\ref{FIG2}.(c) show that the accelerations of follower vehicles and velocity differences between adjacent vehicles converge to zero as the acceleration of leader vehicle goes to zero, respectively. From Fig.\ref{FIG2}.(d), it is observed that the inter-vehicle distance errors converge to zero and they are not amplified in the backward propagation along the platoon.

\begin{figure}[H]
	\centering
	\begin{minipage}[t]{0.236\textwidth}
		\centering
		\includegraphics[width=4.5cm]{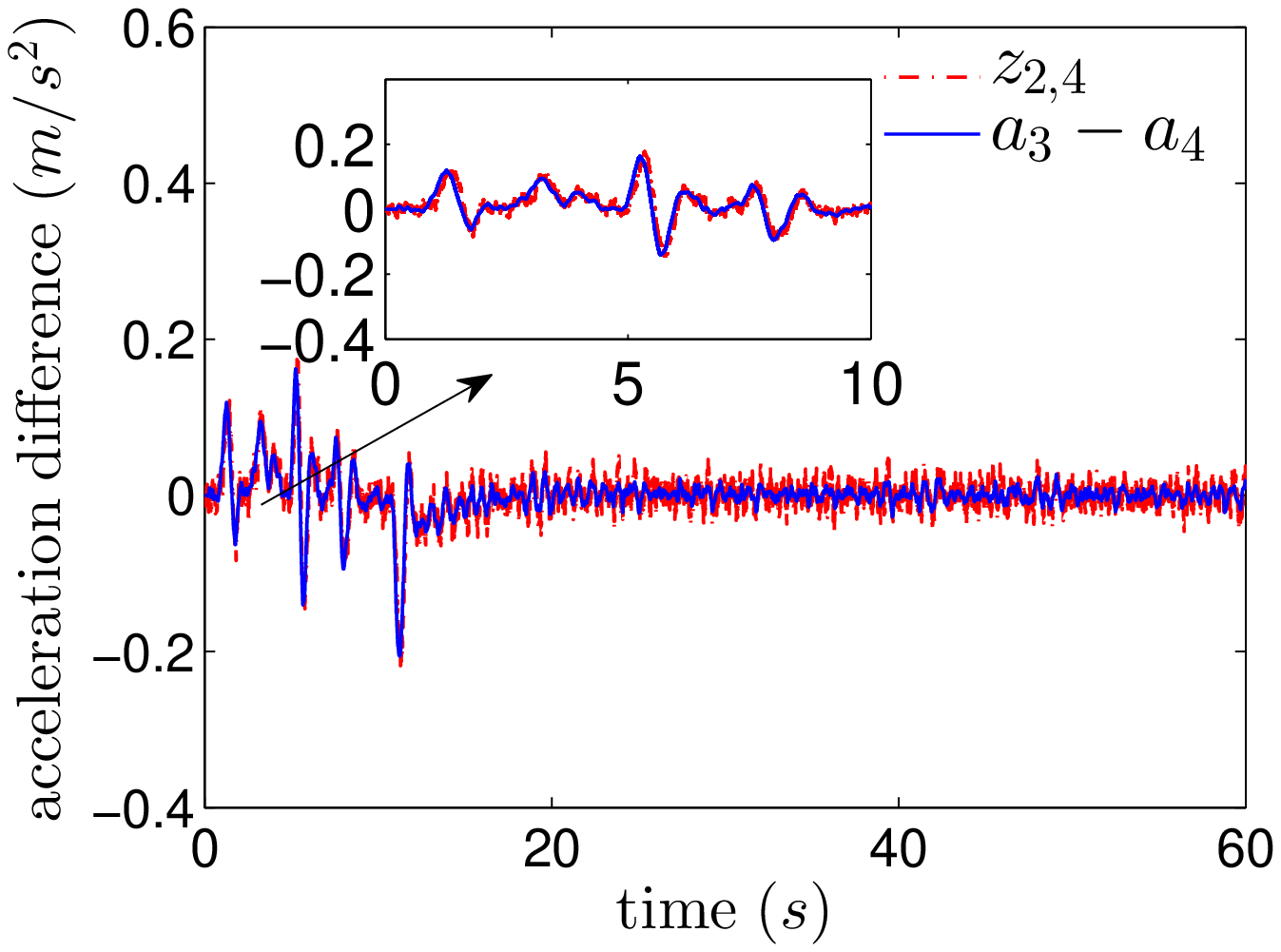}
		\caption{(a)}
\label{FIG3}
	\end{minipage}
	\begin{minipage}[t]{0.236\textwidth}
		\centering
		\includegraphics[width=4.5cm]{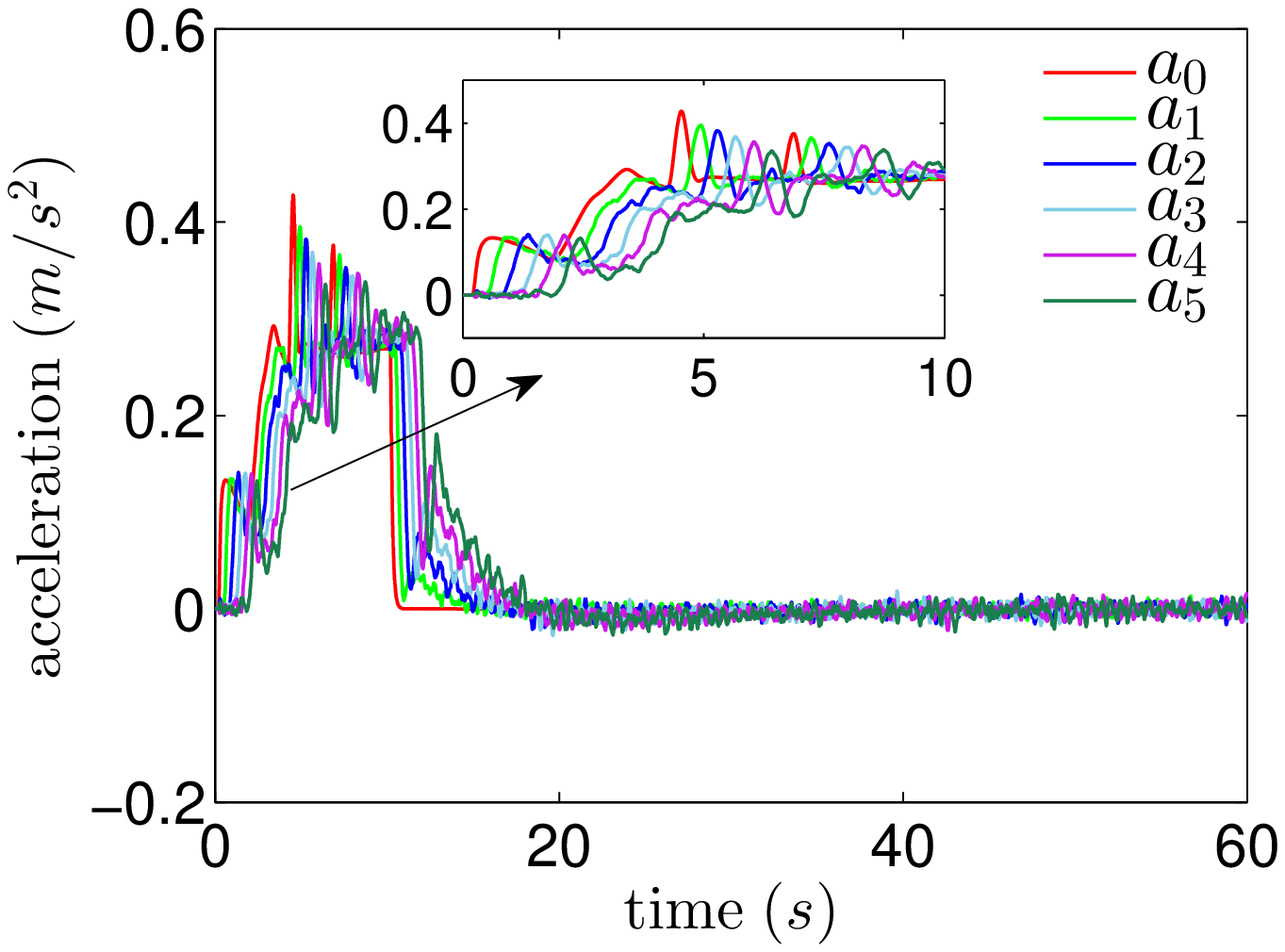}
		\caption*{(b)}
	\end{minipage}
	\centering
	\begin{minipage}[t]{0.236\textwidth}
		\centering
		\includegraphics[width=4.5cm]{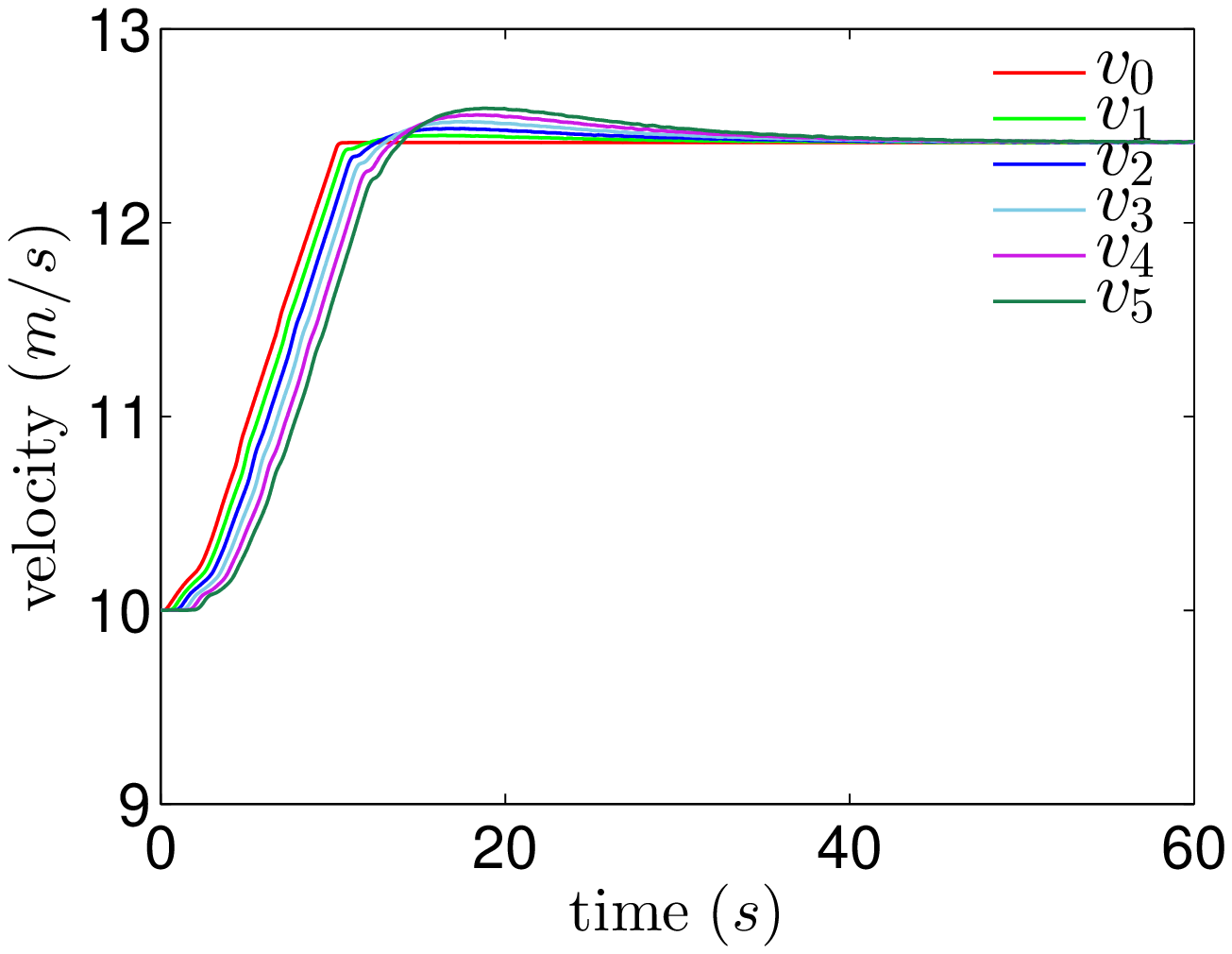}
		\caption*{(c)}
	\end{minipage}
	\begin{minipage}[t]{0.236\textwidth}
		\centering
		\includegraphics[width=4.5cm]{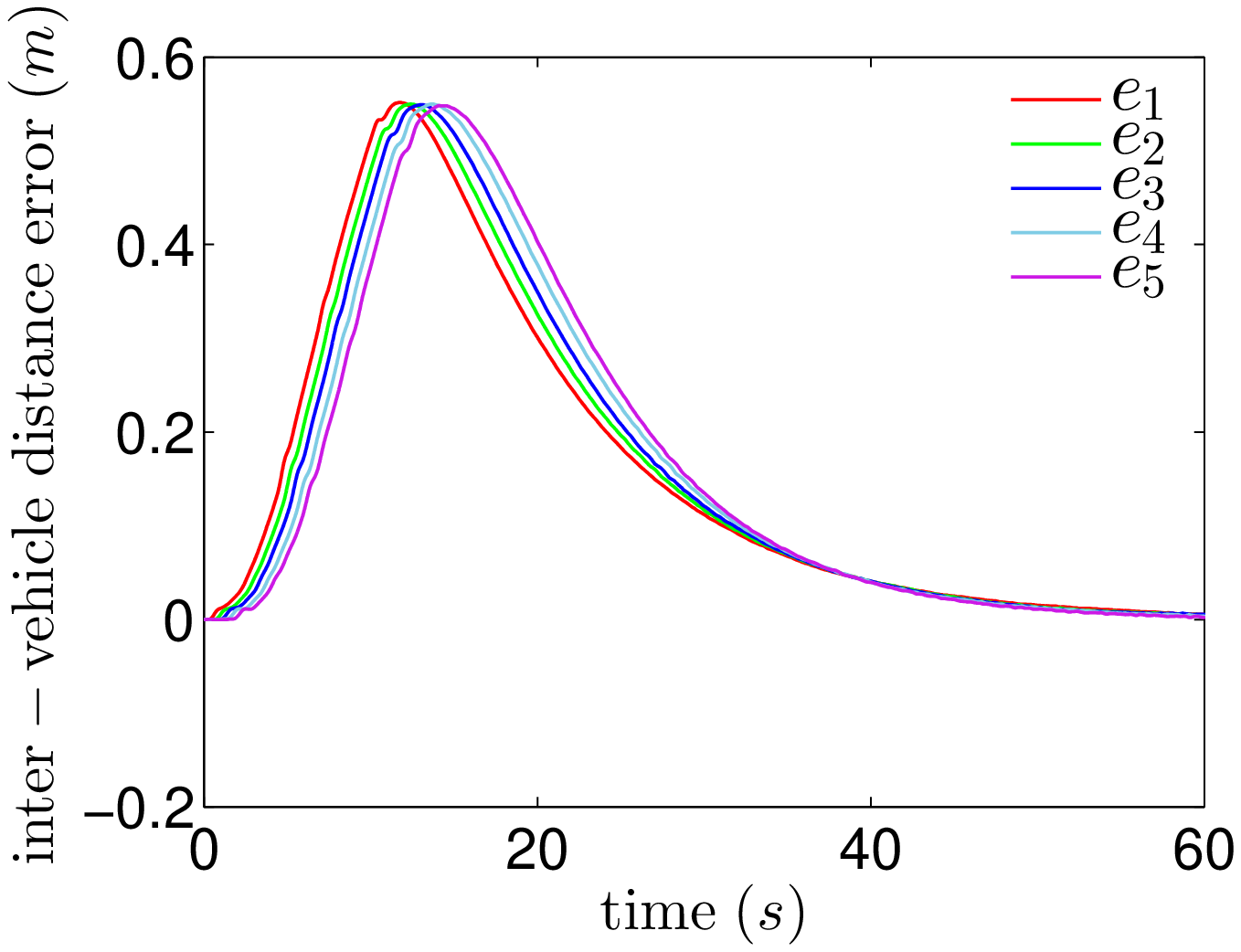}
		\caption*{(d)}
	\end{minipage}
	\caption* {Vehicle model with parameter uncertainties, sensing noises and input delays. (a) The actual and the estimated acceleration differences between the 3rd and 4th follower vehicles. (b) Accelerations of the vehicles. (c)Velocities of the vehicles.  (d) Inter-vehicle distance errors.}
\end{figure}

Next, we consider the vehicle model with input delays, i.e. the $u_i(t)$ is replaced by $u_i(t-\phi)$, where $\phi=0.2$ is the input delay. In practical applications, the velocity differences between adjacent vehicles $v_{d,i}(t)$, $i=1,...,N$, measured by on-board sensors are usually corrupted by random noises. In the numerical simulations, we implement the sampled-data version of the control law (\ref{z_i}) and (\ref{u_i1}) with $v_{d,i}(k\sigma)$ replaced by $v_{d,i}(k\sigma)+\xi_i^d(k\sigma)$, where $\sigma=0.002s$ is the sampling period and $\{\xi_i^d(k\sigma),k=0,1,...\}$ is a sequence of random variables with the uniform distribution $U(-0.005, 0.005)$. We choose $k_p=0.05$, $k_v=0.6$, $k_a=0.8$ and $\omega_o=10$,  $\beta_1=30$, $\beta_2 = 300$ and $\beta_3=1000$. The actual and the estimated acceleration differences between the 3rd and 4th follower vehicles are shown in Fig.\ref{FIG3}.(a). The evolution of vehicles' accelerations, velocities and inter-vehicle distance errors are shown in Fig.\ref{FIG3}.(b), Fig.\ref{FIG3}.(c) and Fig.\ref{FIG3}.(d), respectively.

From Fig.\ref{FIG3}.(a), it is shown that although the velocity differences $v_{d,i}(t)$ are corrupted by sensing noises, the sensing noises are not significantly amplified and the output of the ESO $z_{2i}(t)$ can still track the acceleration differences between adjacent vehicles $a_{d,i}(t)$. Fig.\ref{FIG3}.(b), Fig.\ref{FIG3}.(c) and Fig.\ref{FIG3}.(d) show that the accelerations of follower vehicles, velocity differences between adjacent vehicles and inter-vehicle distance errors converge to a small neighborhood of zero. Fig.\ref{FIG3} shows the robustness of the proposed control law against parameter uncertainties, sensing noises and input delays.

\begin{figure}[H]
	\centering
	\begin{minipage}[t]{0.236\textwidth}
		\centering
		\includegraphics[width=4.5cm]{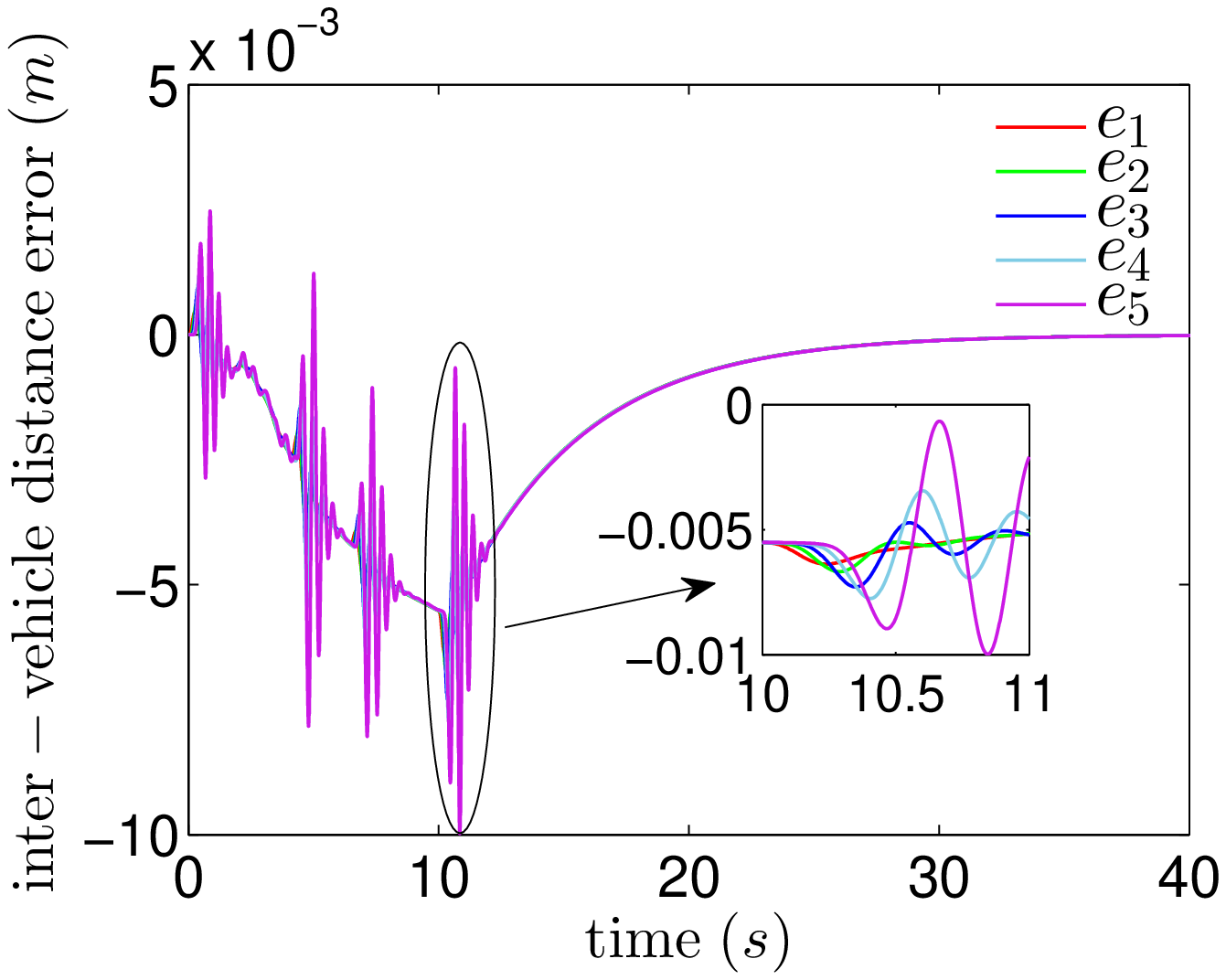}
        \caption{(a)}
        \label{FIG4}
	\end{minipage}
	\begin{minipage}[t]{0.236\textwidth}
		\centering
		\includegraphics[width=4.5cm]{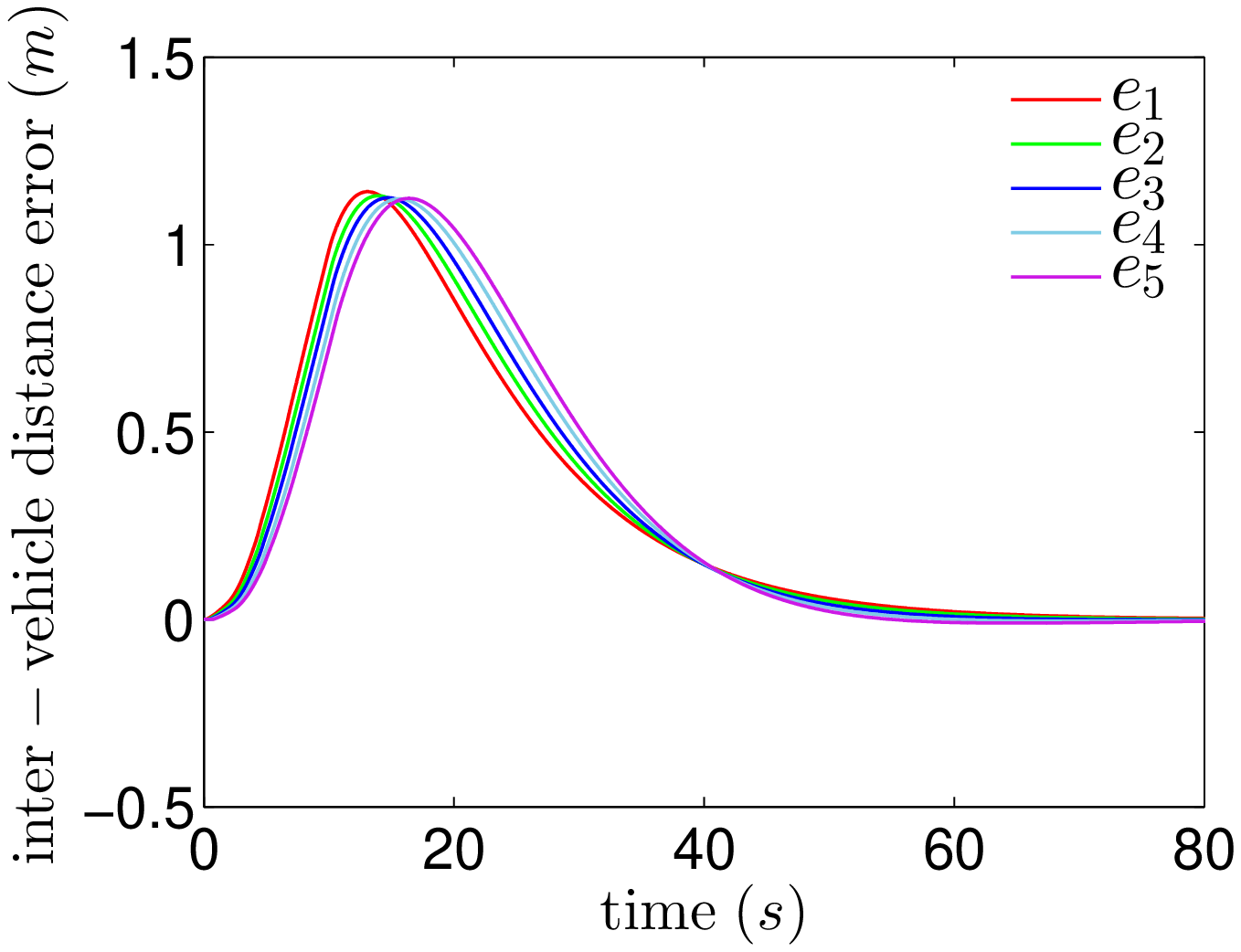}
        \caption*{(b)}
	\end{minipage}
	\caption* {Vehicle platoon under the control law (\ref{z_i}) and (\ref{u_i1}) with $h = 0.01s$. (a) $k_p=8$, $k_v=40$, $k_a=1.2$, $\beta_1=45$, $\beta_2=675$ and $\beta_3=3375$. (b) $k_p=0.01$, $k_v=0.2$, $k_a=0.8$, $\beta_1=45$, $\beta_2=675$ and $\beta_3=3375$. }
\end{figure}

Then let the time headway $h$ change to $0.01s$.  By Fig.\ref{FIG4}.(a), it is observed that the control law (\ref{z_i}) and (\ref{u_i1}) with the previous parameters, i.e. $k_p=8$, $k_v=40$, $k_a=1.2$, $\beta_1=45$, $\beta_2 = 675$ and $\beta_3=3375$, cannot ensure the $\mathcal{L}_2$ string stability any longer, but the closed-loop stability is still guaranteed.

 We reselect $k_p=0.01$, $k_v=0.2$, $k_a=0.8$, $\beta_1=45$, $\beta_2=675$ and $\beta_3=3375$. It can be seen from Fig.\ref{FIG4}.(b) that the closed-loop and $\mathcal{L}_2$ string stabilities are both ensured by reselecting the control parameters. It is shown that for a smaller time headway $h$, a smaller $k_p$ can be chosen to ensure string stability at the cost of slower convergence.

The transient performance of the closed-loop system can be investigated from the structure of the compound controller (\ref{u_i1}) containing proportional differential feedback and feedforward terms. The convergence rate of the closed-loop system is mainly determined by the proportional differential term. The feedforward term reduces the influence of disturbances.
\begin{figure}[H]
	\centering
	\begin{minipage}[t]{0.236\textwidth}
		\centering
		\includegraphics[width=4.5cm]{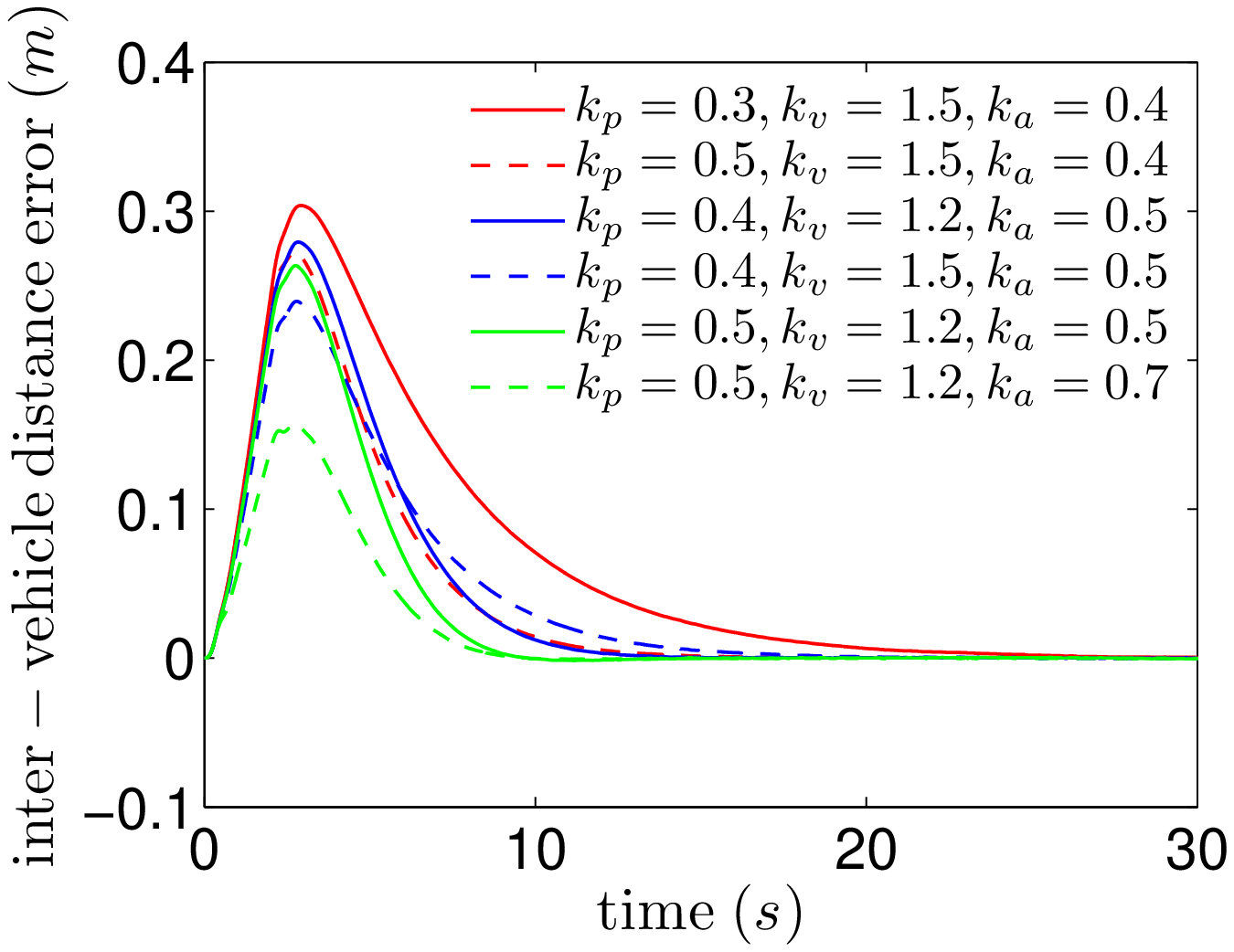}
	\end{minipage}
\caption{The evolution of inter-vehicle distance errors between the leader vehicle and the first follower vehicle with different control parameters. }
\label{FIG5}
\end{figure}

 Let any two of the control parameters $k_p$, $k_v$ and $k_a$  be fixed and the other one be changed. The evolution of inter-vehicle distance errors between the leader and the first follower under different control parameters are shown in Fig.\ref{FIG5}. Fig.\ref{FIG5} shows that a larger $k_p$ leads to a faster convergence and a larger $k_v$ leads to a slower convergence with smaller inter-vehicle distance error. The parameter $k_a$ has little effect on the convergence rate, and a larger $k_a$ brings a smaller inter-vehicle distance error.

\section{Conclusion}
We have considered the platoon control for homogeneous vehicles with third-order linear dynamics model. The constant time headway spacing policy is adopted. Firstly, the distributed cooperative extended state observers are designed to estimate the acceleration differences between adjacent vehicles. Then the controller of each follower vehicle is designed by its own velocity, acceleration, velocity difference and estimated acceleration difference with respect to its immediate predecessor. The information required by the control law can be obtained by on-board sensors. The closed-loop stability of the vehicle platoon system is analyzed by the stability theory of perturbed linear systems, and the sufficient conditions to ensure the closed-loop stability are given. Also the $\mathcal{L}_2$ string stability is analyzed in the frequency domain and the range of the control parameters that guarantee the $\mathcal{L}_2$ string stability is presented. It is shown that for any given positive time headway, control parameters can be properly designed to guarantee both closed-loop and $\mathcal{L}_2$ string stabilities of the vehicle platoon system. In addition, it has been shown that the closed-loop and $\mathcal{L}_2$ string stabilities of the vehicle platoon system can be guaranteed with small model parameter uncertainties.

\section*{Acknowledgements}

This work was supported by the National Natural Science Foundation of China under Grant 61977024 and the Basic
Research Project of Shanghai Science and Technology Commission under Grant 20JC1414000.

\begin{appendices}
\section{}\label{parameters}
Parameter definitions in Theorem \ref{theorem2}
\begin{equation*}
\theta_i=
\left\{
\begin{array}{ll}
\begin{aligned}(\sqrt{\gamma_i^2-4\alpha_i\rho_i}-\gamma_i)/(2\alpha_i),\,{\rm if} \ &\gamma_i^2-4\alpha_i\rho_i\geq0,  \vspace{1ex}  \end{aligned}\\
\begin{aligned}    0, \hspace{38.5mm} {\rm if} \ &\gamma_i^2-4\alpha_i\rho_i<0,\end{aligned}\\
 \hspace{55mm}i = 1,2,3,4,
\end{array}
\right.
\end{equation*}	
\vspace{-8mm}
\begin{align*}
    \rho_1 = &3\tau^2\omega_0^2+1, \rho_2 = 3\tau^2\omega_0^4+3\omega_o^2,\\
    \rho_3 = &\tau^2\omega_o^6+3\omega_o^4, \rho_4 =\omega_o^6, \\
    \alpha_1 = &h^2\mu_v^2, \alpha_2 = 3h^2\mu_v^2\omega_o^2+h^2\mu_p^2,\\
    \alpha_3 = &(3h^2\mu_v^2-9\mu_a^2)\omega_o^4-16\mu_a\mu_v\omega_o^3+(3h^2\mu_p^2-6\mu_p\mu_a)\omega_o^2,  \\
    \alpha_4 = &(h^2\mu_v^2-\mu_a^2)\omega_o^6+(3h^2\mu_p^2+12\mu_a\mu_p)\omega_o^4,\\
    \alpha_5 =&(h^2\mu_p^2+2\mu_a\mu_p)\omega_o^6,\\
    \gamma_1 = & 2(h-\tau)\mu_v-2h\tau\mu_p,\\
    \gamma_2 = &[6(h-\tau)\mu_v-6h\tau\mu_p]\omega_o^2-2\mu_p,\\
    \gamma_3 = &6(h\mu_v-\tau\mu_v-h\tau\mu_p)\omega_o^4-6\mu_p\omega_o^2,  \\
    \gamma_4 =& 2(h\mu_v-\tau\mu_v-h\tau\mu_p)\omega_o^6-6\mu_p\omega_o^4,\gamma_5 = 2\mu_p\omega_o^6.
\end{align*}
Parameter definitions in Theorem \ref{theorem3}
    \begin{align*}
      \Theta_1 = &(k_vh+\tau\beta_2+4\tau+1)/\tau^2, A_1 = \overline \epsilon^2+\overline \epsilon/\tau,\\
       A_2 = &(2N-5)/\tau^2+(4N-8)\overline \epsilon^2+(5N-11)\overline \epsilon/\tau,\\
       Y_1 = &k_vh+1, Y_2 = k_vh/\tau+\beta_2+2,\\
       Y_3 = &k_p+k_v+5k_vh+5,\\
       Y_4 = &(k_p+k_v+6k_vh+3)/\tau+3\beta_2+6,\\
       Y_5 = &(2N-3)(k_p+k_v+2k_vh+2)+(2N-1)(k_vh+1),\\
       Y_6 = &N(k_vh/\tau+\beta_2+2)+(N-1)((k_p+k_v+4k_vh+3)\\
             &/\tau+\beta_2+2)+(N-2)(4+2k_p+2k_v+4k_vh)/\tau,
                     \end{align*}
                     \vspace{-12mm}
                   \begin{align*}
        Z_1(k_p,k_v) = &k_v^2h^2+(2k_v+k_pk_v+k_v^2)h+k_p+k_v+1,\\
        Z_2(k_p,k_v) = &k_v^2h^2/\tau+(k_v^2+k_pk_v+k_v)h/\tau+k_ph+k_vh\\
                       &+2k_p+2k_v+1,\\
        Z_3(k_p,k_v) = &3k_v^2h^2+(3k_v^2+3k_pk_v+6k_v)h+3k_p\\
              &+3k_v+3,\\
        Z_4(k_p,k_v) = &4k_v^2h^2/\tau+((4k_v^2+6k_v+4k_pk_v)h+2k_p\\
        &+2k_v)/\tau+(3k_p+2k_v)h+5k_p+6k_v+4,\\
        Z_5(k_p,k_v) = &(2N-1)(k_pk_vh+k_p+k_v+k_v^2h+1\\
                        &+2k_vh+k_v^2h^2),\\
        Z_6(k_p,k_v) = &N((k_pk_vh+k_v^2h+k_vh+k_v^2h^2)/\tau+1+k_vh\\
                &+2k_p+2k_v+k_ph)+(N-1)((2k_v+2k_p\\
                &+2k_v^2h+2k_pk_vh+2k_v^2h^2+4k_vh)/\tau+k_p\\
                &+k_ph+2k_v+2)+(N-2)k_v/\tau.
    \end{align*}
Parameter definitions in Theorem \ref{theorem4}
\begin{align*}
      \overline b =& \frac{1}{\tau}+\overline \epsilon, \underline b = \frac{1}{\tau}-\overline \epsilon,\overline \rho_1 = 3\omega_o^2+\overline b^2,\\
      \overline \rho_2 =& 3\omega_o^4+3\overline b^2\omega_o^2,\overline \rho_3 = \omega_o^6+3\overline b^2\omega_o^4,\overline \rho_4 =\overline b^2\omega_o^6,\\
\overline \lambda_1 = &3\mu_vh(\overline b^2h\mu_v-2\underline b\mu_a/\tau)+3h^2\mu_a(2\overline b^2\mu_vh-3\underline b^2\mu_a),\\
\underline \lambda_1 = &3\mu_vh(\underline b^2h\mu_v-2\overline b\mu_a/\tau)+3h^2\mu_a(2\underline b^2\mu_vh-3\overline b^2\mu_a),\\
\overline \lambda_2= &16\overline b^2h\mu_a\mu_p-16(\underline b\mu_a\mu_v+\underline bh\mu_a\mu_p)/\tau,\\
\underline \lambda_2= &16\underline b^2h\mu_a\mu_p-16(\overline b\mu_a\mu_v+\overline bh\mu_a\mu_p)/\tau,\\
\overline \lambda_3= &3\overline b^2h^2\mu_p^2+6\overline b^2\mu_a\mu_p-12\underline b\mu_p\mu_a/\tau,\\
\underline \lambda_3= &3\underline b^2h^2\mu_p^2+6\underline b^2\mu_a\mu_p-12\overline b\mu_p\mu_a/\tau,\\
   \overline \alpha_1 = &(\mu_a/\tau-\underline b\mu_a+\overline bh\mu_v)^2,\underline \alpha_1 = (\mu_a/\tau-\overline b\mu_a+\underline bh\mu_v)^2,\\
   \overline \alpha_2 =&((12\overline bh\mu_a\mu_v-18\underline b\mu_a^2)/\tau-12\underline b^2h\mu_a\mu_v+3\overline b^2h^2\mu_v^2\\
             &+9\mu_a^2/\tau^2+9\overline b^2\mu_a^2)\omega_o^2+\overline b^2h^2\mu_p^2+2\overline b^2h \mu_a\mu_p/\tau\\
              &-2\underline b\mu_a\mu_p/\tau,\\
   \underline \alpha_2 =&((12\underline bh\mu_a\mu_v-18\overline b\mu_a^2)/\tau-12\overline b^2h\mu_a\mu_v+3\underline b^2h^2\mu_v^2\\
              &+9\mu_a^2/\tau^2+9\underline b^2\mu_a^2)\omega_o^2+\underline b^2h^2\mu_p^2+2\underline b^2 \mu_a\mu_p/\tau\\
              &-2\overline b\mu_a\mu_p/\tau,\\
    \overline \alpha_3 = &\overline \lambda_1\omega_o^4+\overline \lambda_2\omega_o^3+\overline \lambda_3\omega_o^2,\underline \alpha_3 = \underline \lambda_1\omega_o^4+\underline \lambda_2\omega_o^3+\underline \lambda_3\omega_o^2,\\
    \overline \alpha_4 = &(\overline b^2h^2\mu^2_v-\underline b^2\mu^2_a)\omega_o^6+(3\overline b^2h^2\mu_p^2+6\overline b^2\mu_p\mu_a\\
       &+6\overline b\mu_p\mu_a/\tau)\omega_o^4,
                                                   \end{align*}
\begin{align*}
     \underline \alpha_4 = &(\underline b^2h^2\mu^2_v-\overline b^2\mu^2_a)\omega_o^6+(3\underline b^2h^2\mu_p^2+6\underline b^2\mu_p\mu_a\\
     &+6\underline b\mu_p\mu_a/\tau)\omega_o^4\\
     \overline \alpha_5 =& (\overline b^2h^2\mu_p^2+2\overline b^2\mu_a\mu_p)\omega_o^6,\underline \gamma_5 =2\underline b^2\mu_p\omega_o^6,\\
    \overline \gamma_1 = &2\overline b\mu_a/\tau+2\overline b^2h\mu_v-2\underline b^2\mu_a-2\underline bh\mu_p-2\underline b\mu_v,\\
 \underline \gamma_1 = &2\underline b\mu_a/\tau+2\underline b^2h\mu_v-2\overline b^2\mu_a-2\overline bh\mu_p-2\overline b\mu_v,\\
  \overline \gamma_2 =&16(\mu_a/\tau-\underline b\mu_a)\omega_o^3+(12\overline b\mu_a/\tau+6\overline b^2h\mu_v-12\underline b^2\mu_a\\
    &-6\underline b\mu_v-6\underline bh\mu_p)\omega_o^2-2\underline b^2\mu_p,\\
    \underline \gamma_2 =&16(\mu_a/\tau-\overline b\mu_a)\omega_o^3+(12\underline b\mu_a/\tau+6\underline b^2h\mu_v-12\overline b^2\mu_a\\
    &-6\overline b\mu_v-6\overline bh\mu_p)\omega_o^2-2\overline b^2\mu_p,\\
  \overline \gamma_3 = &(6\overline b^2\mu_a-6\underline b\mu_v-6\underline bh\mu_p-6\underline b\mu_a/\tau+6\overline b^2h\mu_v)\omega_o^4\\
               &+16\overline b^2\mu_a\omega_o^3/\tau-6\underline b^2\mu_p\omega_o^2, \\
  \underline \gamma_3 = &(6\underline b^2\mu_a-6\overline b\mu_v-6\overline bh\mu_p-6\overline b\mu_a/\tau+6\underline b^2h\mu_v)\omega_o^4\\
               &+16\underline b^2\mu_a\omega_o^3/\tau-6\overline b^2\mu_p\omega_o^2,\\
    \overline \gamma_4 =& (2\overline b^2h\mu_v-2\underline bh\mu_p-2\underline b\mu_v)\omega_o^6-6\underline b^2\mu_p\omega_o^4,  \\
    \underline \gamma_4 =& (2\underline b^2h\mu_v-2\overline bh\mu_p-2\overline b\mu_v)\omega_o^6-6\overline b^2\mu_p\omega_o^4,\\
   \theta_{i}=&(\sqrt{(\max\{|\overline \gamma_i|,|\underline \gamma_i|\})^2+4\overline\alpha_i\overline\rho_i}+\max\{|\overline \gamma_i|,\\
              &|\underline \gamma_i|\})/(2\underline\alpha_i),\;i = 1,2,3,4.
\end{align*}
\section{Proof of Theorem 1} \label{Theorem 1}   % Each appendix must have a short title.

	The proof of Theorem 1 needs the following lemmas.

\begin{lemma}{\rm \citep{hinrichsen1986stability,guo1993time}}
	Suppose $\dot x(t) = Ax(t)$ is exponentially stable, where $A\in \mathbb{C}^{n\times n}$.  Denote $r_c(A)= \min \limits_{\omega \in \mathbb{R}}S_n(i\omega I-A)$. If  $\|B\|<r_c(A)$, then $\dot x(t) = (A+B)x(t)$ is exponentially stable. Further, there exists $B\in \mathbb{C}^{n\times n}$ with $\|B\|=r_c(A)$, such that $\dot x(t) = (A+B)x(t)$ is not asymptotically stable.
\end{lemma}

\vskip 0.2cm

\begin{lemma}
	For any $A$ $\in$ $\mathbb{R}^{n\times n}$, $\|A\|\leq \sum^n_{i=1} \sum^n_{j=1}|a_{ij}|$, where $a_{ij}$ is the element of the $i$th row and $j$th column of $A$.
\end{lemma}

\textbf{Proof.}  Denote \begin{equation}\label{b_pq}
b_{pq}^{ij}=\left\{
\begin{array}{ll}
a_{ij},  &p=i, q=j, \vspace{1ex} \\
0,  &\rm{otherwise}.
\end{array}
\right.
\end{equation}
Define $A_{ij} = \left[b_{pq}^{ij}\right]_{n\times n}$, where $b_{pq}^{ij}$ is the element of the $p$th row and $q$th column of $A_{ij}$.

By (\ref{b_pq}) and the definition of $A_{ij}$, we have
\begin{align*}%\label{b_pq1}
A = \sum^n_{i=1}\sum^n_{j=1}A_{ij}.
\end{align*}
This together with the triangle inequality of matrix norm leads to
\begin{align}\label{b_pq2}
\|A\|\leq \sum^n_{i=1}\sum^n_{j=1}\|A_{ij}\|.
\end{align}
By the definition of the 2-norm of matrix, we have
\begin{align}\label{fanshu}
\|A_{ij}\| = \sqrt{\lambda_{max}(A_{ij}^TA_{ij})},
\end{align}
where $\lambda_{max}(A_{ij}^TA_{ij})$ is the maximum eigenvalue of $A_{ij}^TA_{ij}$.

By (\ref{b_pq}) and the definition of $A_{ij}$, we get
\begin{align*}%\label{A_{ij}^TA_{ij}}
(A_{ij}^TA_{ij})_{pq}&=\sum^n_{k=1} b_{kp}^{ij}b_{kq}^{ij}
=\left\{
\begin{array}{ll}
a_{ij}^2  ,&p=j, q=j, \vspace{1ex} \\
0  ,&\rm{otherwise}.
\end{array}
\right.
\end{align*}
This together with (\ref{fanshu}) gives $\|A_{ij}\|=|a_{ij}|$. Then by (\ref{b_pq2}), we get $\|A\|\leq \sum^n_{i=1}\sum^n_{j=1}|a_{ij}|$. \qed

\vskip 0.2cm

\textbf{Proof of Theorem 1.} Denote
\begin{align}
X_i(t)=&[p_i(t),v_i(t),a_i(t)]^{T},\;i=0,1,2,...,N\label{x16}, \vspace{1ex}\\
E_i(t)=&[e_{1,i}(t),e_{2,i}(t),e_{3,i}(t)]^{T},\;i=1,2,...,N\label{e16}, \vspace{1ex}\\
F_i(t) = &[e_i(t),v_{d,i}(t),a_i(t)]^{T},\;i=1,2,...N, \label{f18}\vspace{1ex}\\
W(t) = &[F_1^{T}(t),F_2^{T}(t),\cdots,F_N^{T}(t),E_1^{T}(t),E_2^{T}(t), \nonumber \\
&\cdots,E_N^{T}(t)]^{T},\label{W} \vspace{1ex}\\
\Delta(t) = &[\delta_1^{T}(t),0,\cdots,0,\zeta_1^{T}(t),\zeta_2^{T}(t),\zeta_3^{T}(t),0,\nonumber \\
&\cdots,0]^{T},\label{delta}
\end{align}
where
\begin{align}
e_{1,i}(t)=&z_{1,i}(t)-v_{d,i}(t), \label{e_1i}\\
e_{2,i}(t)=&z_{2,i}(t)-a_{d,i}(t), \label{e_2i}\\
e_{3,i}(t)=&z_{3,i}(t)-q_i(t), \label{e_3i}\\
\delta_1(t)=& \left[0,a_0(t),-k_aa_0(t)/\tau\right]^{T}, \label{delta_1}\\
\zeta_1(t) =& \left[0,0,(u_0(t)-a_0(t)-\tau\dot u_0(t))/\tau^2\right]^{T},\label{zeta_1}\\
\zeta_2(t) = & \left[0,0,(k_aa_0(t)-\tau k_va_0(t)-k_ak_vha_0(t)\right. \nonumber \\
&\left.-\tau k_a\dot a_0(t))/\tau^2\right]^{T}, \label{zeta_2}\\
\zeta_3(t)=& \left[0,0,k_a^2a_0(t)/\tau\right]^{T}.\label{zeta_3}
\end{align}

From (\ref{a_di1}), (\ref{u_i1}) and (\ref{e_2i}), we know
\begin{align}\label{u_i2}
u_i(t) = &\;k_pe_i(t)+k_v(v_{d,i}(t)-ha_i(t)) \nonumber\\
&+k_a(a_{i-1}(t)+e_{2i}(t)),\;i=1,2,\ldots,N.
\end{align}
This together with (\ref{x_i}), (\ref{e_i}), (\ref{v_di1}), (\ref{x16}) and (\ref{e16}) leads to
\begin{align}\label{dX_i}
\dot{X_i}(t)=\left\{
\begin{array}{ll}
A_0X_i(t)+B_0u_i(t),&\hspace{-3mm}i=0, \vspace{1ex} \\
AX_i(t)+BX_{i-1}(t)+CE_i(t)&\hspace{-1.0mm}+Lr,\\
&\hspace{-3mm}i=1,\ldots,N,
\end{array}
\right.
\end{align}
where
\begin{equation}
A_0
=\begin{bmatrix}
0 & 1 & 0\\
0 & 0 & 1\\
0 & 0 & \begin{aligned}-\frac{1}{\tau}\end{aligned}
\end{bmatrix} \nonumber,
B_0
= \begin{bmatrix}
0 \\
0  \\
\frac{1}{\tau}
\end{bmatrix}\nonumber,
L
= \begin{bmatrix}
0 \\
0  \\
\begin{aligned}-\frac{k_p}{\tau}\end{aligned}
\end{bmatrix}\nonumber,
\end{equation}
\begin{equation}
A
=\begin{bmatrix}
0 & 1 & 0\\
0 & 0 & 1\\
\begin{aligned}-\frac{k_p}{\tau}\end{aligned} & \begin{aligned}-\frac{k_v+k_ph}{\tau}\end{aligned} & \begin{aligned}-\frac{1+k_vh}{\tau}\end{aligned}
\end{bmatrix},  \nonumber
\end{equation}
\begin{equation}
B
= \begin{bmatrix}
0 & 0 & 0\\
0 & 0 & 0 \\
\begin{aligned}\frac{k_p}{\tau}\end{aligned} & \begin{aligned}\frac{k_v}{\tau}\end{aligned} & \begin{aligned}\frac{k_a}{\tau}\end{aligned}
\end{bmatrix} \nonumber,
C
=\begin{bmatrix}
0 & 0 & 0 \\
0 & 0 & 0 \\
0 & \begin{aligned}\frac{k_a}{\tau}\end{aligned} & 0
\end{bmatrix} \nonumber.
\end{equation}

From (\ref{e_i}), (\ref{x16}) and (\ref{f18}), we get
\begin{eqnarray}\label{F_i}
F_i(t)=PX_{i-1}(t) - QX_i(t)-L_1r,\;i=1,2,\ldots,N,
\end{eqnarray}
where
\begin{equation}
P = {
	\left[ \begin{array}{ccc}
	1 & 0 & 0\\
	0 & 1 & 0\\
	0 & 0 & 0
	\end{array}
	\right ]} \nonumber,
Q = {
	\left[ \begin{array}{ccc}
	1 & h & 0\\
	0 & 1 & 0\\
	0 & 0 & -1
	\end{array}
	\right ]} \nonumber,
L_1 = {
	\left[ \begin{array}{ccc}
	1 \\
	0 \\
	0
	\end{array}
	\right ]}. \nonumber
\end{equation}
From (\ref{delta_1}), (\ref{dX_i}) and (\ref{F_i}), we obtain
\begin{align}\label{dF_i}
\dot{F_i}(t)=\left\{
\begin{array}{ll}
\mathcal{A} F_i(t)+\mathcal{G}E_i(t)+\delta_i(t),&\hspace{-0.5mm}i = 1, \vspace{1ex} \\
\mathcal{A} F_i(t)+\mathcal{B}_1F_{i-1}(t)+\mathcal{G}E_i(t),&\hspace{-0.5mm}i=2,\ldots,N,
\end{array}
\right.
\end{align}
where
\begin{equation}
\mathcal{B}_1
= \begin{bmatrix}
0 & 0 & 0\\
0 & 0 & 1 \\
0 & 0 & \begin{aligned}\frac{k_a}{\tau}\end{aligned}
\end{bmatrix} \nonumber,
\mathcal{G}
= \begin{bmatrix}
0 & 0 & 0\\
0 & 0 & 0 \\
0 & \begin{aligned}\frac{k_a}{\tau}\end{aligned} & 0
\end{bmatrix}.\nonumber
\end{equation}
By (\ref{e_i})-(\ref{z_i}), (\ref{e16}), (\ref{f18}), (\ref{e_1i})-(\ref{u_i2}) and (\ref{dF_i}), we have
\begin{align}\label{dE_i}
\dot{E_i}(t)=\left\{
\begin{array}{ll}
\mathcal{H}E_i(t)+\zeta_i(t) , &\hspace{-9mm}i = 1, \vspace{1ex}\\
\mathcal{D}F_{i-1}(t)+\mathcal{I}E_{i-1}(t)+\zeta_i(t) , &\hspace{-9mm}i = 2, \vspace{1ex}\\
\mathcal{D} F_{i-1}(t)+\mathcal{E}_1 F_{i-2}(t)+\mathcal{I}E_{i-1}(t)\\
+\mathcal{J}E_{i-2}(t)+\zeta_i(t) , &\hspace{-9mm}i = 3,\vspace{1ex}\\
\mathcal{D} F_{i-1}(t)+\mathcal{E}_1 F_{i-2}(t)+\mathcal{F}F_{i-3}(t)\\
+\mathcal{I}E_{i-1}(t)+\mathcal{J}E_{i-2}(t),  &\hspace{-9mm}i=4,\ldots,N,
\end{array}
\right.
\end{align}
where
\begin{equation}
\mathcal{J}
= \begin{bmatrix}
0 & 0 & 0 \\
0 & 0 & 0 \\
0 & \begin{aligned}-\frac{k_a^2}{\tau^2}\end{aligned} & 0
\end{bmatrix}, \nonumber
\mathcal{F}
=\begin{bmatrix}
0 & 0 & 0\\
0 & 0 & 0\\
0 & 0 & \begin{aligned}-\frac{k_a^2}{\tau^2}\end{aligned}
\end{bmatrix}, \nonumber
\end{equation}
\begin{equation}
\mathcal{I}
= \begin{bmatrix}
0 & 0 & 0\\
0 & 0 & 0 \\
\begin{aligned}\frac{\beta_2k_a}{\tau}\end{aligned} & \begin{aligned}\frac{(1+k_vh)k_a}{\tau^2}\end{aligned} & \begin{aligned}-\frac{k_a}{\tau}\end{aligned}
\end{bmatrix}. \nonumber
\end{equation}
\begin{equation}
\mathcal{E}_1
= \begin{bmatrix}
0 & 0 & 0\\
0 & 0 & 0 \\
\begin{aligned}-\frac{k_pk_a}{\tau^2}\end{aligned} & \begin{aligned}-\frac{k_vk_a}{\tau^2}\end{aligned} & \begin{aligned}-\frac{k_v}{\tau}+\frac{2(1+k_vh)k_a}{\tau^2}\end{aligned}
\end{bmatrix}, \nonumber
\end{equation}
From (\ref{W}), (\ref{delta}), (\ref{dF_i}) and (\ref{dE_i}), we know
\begin{align}\label{closed-loop-2}
\dot{W}(t) = (\Psi+ \hat \Psi)W(t)+\Delta(t),
\end{align}
where
$\hat \Psi
=\begin{bmatrix}
\hat \Psi_{11} & \hat \Psi_{12}  \\
\hat \Psi_{21} & \hat \Psi_{22}
\end{bmatrix}$ and
\begin{equation}
\hat \Psi_{11}
=\begin{bmatrix}
O & \quad & \quad & \quad   \\
\mathcal {B}_2 & O & \quad & \quad  \\
\quad & \ddots & \ddots & \quad  \\
\quad & \quad  & \mathcal {B}_2 & O \\
\end{bmatrix},\nonumber
\hat \Psi_{22}
=\begin{bmatrix}
O & \quad & \quad & \quad & \quad \\
\mathcal {I} & O& \quad & \quad & \quad \\
\mathcal {J} & \mathcal {I} & O & \quad & \quad \\
\quad & \ddots & \ddots & \ddots & \quad \\
\quad & \quad & \mathcal {J}  & \mathcal {I} & O
\end{bmatrix}, \nonumber
\end{equation}
\begin{equation}
\hat \Psi_{21}
=\begin{bmatrix}
O& \quad & \quad & \quad & \quad & \quad \\
O & O & \quad & \quad & \quad & \quad \\
\mathcal {E}_2 & O & O & \quad & \quad & \quad \\
\mathcal {F} & \mathcal {E}_2 & O & O & \quad & \quad \\
\quad & \ddots & \ddots & \ddots & \ddots & \quad \\
\quad  & \quad & \mathcal {F} & \mathcal {E}_2 & O & O
\end{bmatrix},\nonumber
\hat \Psi_{12}
=\begin{bmatrix}
\mathcal {G} & \quad & \quad  \\
\quad & \ddots & \quad  \\
\quad & \quad  &  \mathcal {G} \\
\end{bmatrix},\nonumber
\end{equation}
\begin{align}\nonumber
\mathcal {B}_2=\mathcal {B}_1-\mathcal {B},\mathcal {E}_2=\mathcal {E}_1-\mathcal {E}.
\end{align}
Firstly, we analyze the stability of $\Psi$. The eigenvalues of $\Psi$ are only related to $\mathcal{A}$ and $\mathcal{H}$. Calculating the characteristic polynomial of $\mathcal{A}$, we obtain
\begin{align}\label{A_d}
\left|sI- \mathcal{A} \right|= s^3+\left(\frac{1+k_vh}{\tau}\right)s^2+\left(\frac{k_v+k_ph}{\tau}\right)s+\frac{k_p}{\tau}.
\end{align}
The Rouse table corresponding to (\ref{A_d}) is given by
\begin{table}[H]
	\centering
	\begin{tabular}{c c c }
		$s^3$ & 1 & $\begin{aligned}\frac{k_v+k_ph}{\tau}\end{aligned}$ \\
		$s^2$ & $\begin{aligned}\frac{1+k_vh}{\tau}\end{aligned}$ & $\begin{aligned}\frac{k_p}{\tau}\end{aligned}$  \\
		$s^1$ & $\begin{aligned}\frac{hk_v^2+(1+h^2k_p)k_v+(h-\tau)k_p}{\tau+\tau k_vh}\end{aligned}$ & 0  \\
		$s^0$ & $\begin{aligned}\frac{k_p}{\tau}\end{aligned}$ &  \\
	\end{tabular}
\end{table}
By $k_p>0$ and (\ref{h3}), we know that the elements of the first column of the Rouse table corresponding to (\ref{A_d}) are all greater than zero. From Rouse criterion, $\mathcal{A}$ is stable.
Calculating the characteristic polynomial of $\mathcal{H}$, we get
\begin{align}\label{S}
\left|sI- \mathcal{H}\right| = s^3+\beta_1s^2+\beta_2s+\beta_3.
\end{align}
The Rouse table corresponding to (\ref{S}) is given by
\begin{table}[H]
	\centering
	\begin{tabular}{c c c }
		$s^3$ & 1 & $\beta_2$ \\
		$s^2$ & $\beta_1$ & $\beta_3$  \\
		$s^1$ & $\begin{aligned}\frac{\beta_1 \beta_2-\beta_3}{\beta_1}\end{aligned}$ & 0  \\
		$s^0$ & $\beta_3$ &  \\
	\end{tabular}
\end{table}
By $\beta_1>0$, $\beta_3>0$, $\beta_1\beta_2-\beta_3>0$, we know that the elements of the first column of the Rouse table corresponding to (\ref{S}) are all greater than zero. From Rouse criterion, $\mathcal{H}$ is stable. Then $\Psi$ is stable.
From the definition of $\|\hat{\Psi}\|$ and Lemma 2, we know
\begin{align}\label{hatpsi}
\|\hat{\Psi}\|\leq\left\{
\begin{array}{lc}
\begin{aligned}k_a/\tau\end{aligned},\;&$\rm {if}$ \ N=1, \vspace{1ex}\\
\begin{aligned}k_a(k_vh+\tau\beta_2+4\tau+1)/\tau^2\end{aligned},\;&$\rm {if}$ \ N=2, \vspace{1ex}\\
\begin{aligned}(2N-5)k_a^2/\tau^2+\Theta k_a\end{aligned},\;&$\rm{if}$ \ N\geq3.
\end{array}
\right.
\end{align}
From (\ref{k_a}) and (\ref{hatpsi}), we get $\|\hat{\Psi}\|<r_c(\Psi)$. It is known from the definition of $\Delta(t)$ and Assumption 1 that $\lim \limits_ {t\to \infty}\Delta(t)=0$. By Lemma 1 and (\ref{closed-loop-2}), we know $W(t)$ converges to zero exponentially. Then $v_{i-1}(t)-v_i(t)$ and $e_i(t)$ both converge to zero exponentially, $i=1,2,...,N$, which implies $v_{i}(t)-v_0(t)$ converges to zero exponentially, $i=1,2,...,N$.  \qed

\section{Proof of Theorem 2}\label{Theorem 2}

\textbf{Proof of Theorem 2.} By (\ref{e_i}) and (\ref{v_di1}), we get
\begin{align}\label{a_is}
a_i(t) = \frac{v_{d,i}(t)- \dot e_i(t)}{h}.
\end{align}
Taking the Laplace transform of (\ref{a_is}), we have
\begin{align}\label{La_is}
\mathscr{A}_i(s) = \frac{\mathscr{V}_{d,i}(s)- s\mathscr{E}_i(s)}{h},
\end{align}
where $\mathscr{A}_i(s)$ and $\mathscr{V}_{d,i}(s)$ are the Laplace transform of $a_i(t)$, $v_{d,i}(t)$, respectively.
From (\ref{x_i}), we know
\begin{align}\label{u_i3}
u_i(t)=\tau \dot a_i(t) + a_i(t).
\end{align}
This together with (\ref{a_is}) leads to
\begin{align}\label{u_is}
u_i(t)=\tau \frac{\dot v_{d,i}(t)- \ddot e_i(t)}{h}+\frac{v_{d,i}(t)- \dot e_i(t)}{h}.
\end{align}
Taking the Laplace transform of (\ref{u_is}), we get
\begin{align}\label{Lu_is}
\mathscr{U}_i(s)=\tau \frac{s\mathscr{V}_{d,i}(s)- s^2\mathscr{E}_i(s)}{h}+\frac{\mathscr{V}_{d,i}(s)- s\mathscr{E}_i(s)}{h},
\end{align}
where $\mathscr{U}_i(s)$ is the Laplace transform of $u_i(t)$.
Taking the Laplace transform of (\ref{z_i}), we have
\begin{align}\label{z_is}
\left\{
\begin{array}{lll} %Éè¶šÁÐÕóµÄžñÊœ£º{lll} ÊÇÈýžöL£¬±íÊŸÈýÁÐµÄ¶ÔÆë·œÊœÎªLeft ¶ÔÆë
s\mathscr{Z}_{1,i}(s) = &\mathscr{Z}_{2,i}(s) +\beta_{1}(\mathscr{V}_{d,i}(s)-\mathscr{Z}_{1,i}(s)), \vspace{1ex} \\
s\mathscr{Z}_{2,i}(s) = &\mathscr{Z}_{3,i}(s) +\beta_{2}(\mathscr{V}_{d,i}(s)-\mathscr{Z}_{1,i}(s))+\mathscr{A}_i(s)/\tau\\
&- \mathscr{U}_i(s)/\tau, \vspace{1ex}  \\
s\mathscr{Z}_{3,i}(s) =& \beta_{3}(\mathscr{V}_{d,i}(s)-\mathscr{Z}_{1,i}(s)),\vspace{1ex}
\end{array} %·œ³ÌÁÐÕóµÄœáÊø
\right. %·œ³Ì×éµÄÓÒ±ßÎÞ·ûºÅ£¬ÀûÓÃ¡°.¡°ÀŽ±êÊŸ
\end{align}
%\begin{center}
%$i = 1,2,\ldots,N$,
%\end{center}
where $\mathscr{Z}_{1,i}(s)$, $\mathscr{Z}_{2,i}(s)$ and $\mathscr{Z}_{3,i}(s)$ are the Laplace transform of $z_{1,i}(t)$, $z_{2,i}(t)$ and $z_{3,i}(t)$, respectively.
Substituting (\ref{Lu_is}) into (\ref{z_is}), we obtain
\begin{align}\label{z_2is}
\mathscr{Z}_{2,i}(s) =&\; ((-s^3+(h\beta_2-\beta_1)s^2+h\beta_3s)\mathscr{V}_{d,i}(s)+(s^4 \nonumber \\
&+\beta_1s^3)\mathscr{E}_i(s)/(h(s^3+\beta_1s^2+\beta_2s+\beta_3))
\end{align}
By (\ref{u_i1}), (\ref{a_is}) and (\ref{u_i3}), we get
\begin{align}\label{22}
\tau \dot a_i(t)+a_i(t) = &\;k_{p}e_i(t)+k_{v}\dot e_i(t) \nonumber \\
&+k_{a}(z_{2,i}(t)+a_i(t)).
\end{align}
Taking the Laplace transform of (\ref{22}), we have
\begin{align}\label{2}
\tau s\mathscr{A}_i(s)+\mathscr{A}_i(s)=&\;k_p\mathscr{E}_i(s)+k_v s\mathscr{E}_i(s)+k_a\mathscr{A}_i(s)\nonumber\\
&+k_a\mathscr{Z}_{2,i}(s).
\end{align}
Denote $H(s)=\frac{\mathscr{V}_{d,i}(s)}{\mathscr{E}_i(s)}$. By (\ref{La_is}), (\ref{z_2is}) and (\ref{2}), we get
\begin{align}\label{HH}
H(s)=\frac{\tau s^5+ n_4s^4+ n_3s^3+ n_2s^2+ n_1s+ n_0}{\tau s^4+d_3s^3+d_2s^2+d_1s+d_0},
\end{align}
where
\begin{align*}
&n_0 = k_ph\beta_3,  \\
&n_1 = k_ph\beta_2+(1-k_a+k_vh)\beta_3, \\
&n_2 = k_ph\beta_1+(1-k_a+k_vh)\beta_2+\tau \beta_3,\nonumber\\
&n_3 = k_ph+(1+k_vh)\beta_1+\tau\beta_2, \\
&n_4 = 1+k_vh+\tau\beta_1,\\
&d_0 = (1-k_a)\beta_3,\\
&d_1 = (1-k_a)\beta_2+(\tau-k_ah)\beta_3,\nonumber \\
&d_2 = \beta_1+(\tau-k_ah)\beta_2,  \\
&d_3 = \tau \beta_1+1.
\end{align*}
By (\ref{e_i}) and (\ref{v_di1}), we get
\begin{align}\label{de}
\dot e_{i-1}(t)- \dot e_i(t)=v_{d,i-1}(t)-v_{d,i}(t)-h\dot v_{d,i}(t).
\end{align}
 Taking the Laplace transform of (\ref{de}), we have
\begin{align*}%\label{Lde}
s\mathscr{E}_{i-1}(t)- s\mathscr{E}_i(t)=\mathscr{V}_{d,i-1}(t)-\mathscr{V}_{d,i}(t)-hs\mathscr{V}_{d,i}(t).
\end{align*}
This together with $\mathscr{V}_{d,i}(s)=H(s)\mathscr{E}_i(s)$ leads to
\begin{align*}%\label{G_es1}
G_{ei}(s) =\frac{s-H(s)}{s-(hs+1)H(s)}.
\end{align*}
This together with (\ref{HH}) leads to
%\begin{align}\label{G_es}
%&G_e(s) \nonumber\\
% &=\frac{ \overline n_4s^4+\overline n_3s^3+ \overline n_2s^2+ \overline n_1s+ \overline n_0}{\overline d_6s^6+\overline d_5s^5+\overline d_4s^4+\overline d_3s^3+\overline d_2s^2+\overline d_1s+\overline d_0},
%\end{align}
\begin{align}\label{G_es}
G_{ei}(s)= &\;(k_vs^4+\overline n_3s^3+ \overline n_2s^2+ \overline n_1s+ k_p\beta_3)/(\tau s^6 \nonumber \\
&+\overline d_5s^5+\overline d_4s^4+\overline d_3s^3+\overline d_2s^2+\overline d_1s+k_p\beta_3),
\end{align}
where
\begin{align*}
\overline n_1 = &k_p\beta_2+k_v\beta_3,\\
\overline n_2 = &k_p\beta_1+k_v\beta_2+k_a\beta_3, \\
\overline n_3 = &k_v\beta_1+k_a\beta_2+k_p,  \\
\overline d_1 = &k_p\beta_2+(k_ph+k_v)\beta_3,  \\
\overline d_2 = &k_p\beta_1+(k_ph+k_v)\beta_2+(1+k_vh)\beta_3,\\
\overline d_3 = &(k_ph+k_v)\beta_1+(1+k_vh)\beta_2+\tau\beta_3+k_p,\\
\overline d_4 = &(1+k_vh)\beta_1+\tau \beta_2+k_ph+k_v, \\
\overline d_5 = &\tau \beta_1+k_vh+1.
\end{align*}
Substituting $s=j\omega$ into (\ref{G_es}), we get
\begin{align}\label{G_e}
G_{ei}(j\omega)=\frac{x_n(\omega)+y_n(\omega)j}{x_d(\omega)+y_d(\omega)j},
\end{align}
where $x_n(\omega)=k_p\beta_3- \overline n_2\omega^2+ k_v\omega^4$, $y_n(\omega) =\overline n_1\omega- \overline n_3\omega^3 $, $x_d(\omega)=k_p\beta_3- \overline d_2\omega^2+\overline d_4\omega^4-\tau\omega^6$, $y_d(\omega)=\overline d_1\omega-\overline d_3\omega^3+\overline d_5\omega^5$.

By $\mu_p>0$ and $\mu_a>0$, we know $\alpha_5>0$ and $\gamma_5>0$. From (\ref{k}), we obtain $k\geq \gamma_5/\alpha_5$. This together with $\alpha_5>0$ and $\gamma_5>0$ leads to
\begin{align}\label{k5}
\alpha_5k^2-\gamma_5k\geq0.
\end{align}
From (\ref{mu_v}) and $\mu_a>0$, we know $\mu_v>\mu_a/h$. This together with $\mu_p>0$ leads to $\alpha_4>0$. From (\ref{k}), we know $k\geq\theta_4$. This together with $\alpha_4>0$ and $\rho_4>0$ leads to
\begin{align}\label{k4}
\alpha_4k^2+\gamma_4k+\rho_4\geq0.
\end{align}
By (\ref{mu_v}), we know $\mu_v>\sqrt{3}\mu_a/h$. This together with $\mu_a>0$ leads to  $3h^2\mu_v^2-9\mu_a^2>0$. By (\ref{mu_v}), we get $\mu_v>2\mu_a/h^2$. This leads to $3h^2\mu_p^2-6\mu_p\mu_a>0$. By (\ref{omega}), we know $\omega_o>16\mu_v\mu_a/(3h^2\mu_v^2-9\mu_a^2)$. This together with $3h^2\mu_v^2-9\mu_a^2>0$ leads to $\alpha_3>0$. From (\ref{k}), we know $k\geq\theta_3$. This together with $\alpha_3>0$ and $\rho_3>0$ leads to
\begin{align}\label{k3}
\alpha_3k^2+\gamma_3k+\rho_3\geq0.
\end{align}
From (\ref{k}), we obtain $k\geq\theta_2$. This together with $\alpha_2>0$ and $\rho_2>0$ leads to
\begin{align}\label{k2}
\alpha_2k^2+\gamma_2k+\rho_2\geq0.
\end{align}
From (\ref{k}), we obtain $k\geq \theta_1$. This together with $\alpha_1>0$ and $\rho_1>0$ leads to
\begin{align}\label{k1}
\alpha_1k^2+\gamma_1k+\rho_1\geq0.
\end{align}
By (\ref{k5})-(\ref{k1}), we know
\begin{align}\label{equ2}
&(\alpha_5k^2-\gamma_5k)\omega^{2}+(\alpha_4k^2+\gamma_4k+\rho_4)\omega^{4} \nonumber \\
&\hspace{-4mm}+(\alpha_3k^2+\gamma_3k+\rho_3)\omega^{6}+(\alpha_2k^2+\gamma_2k+\rho_2)\omega^{8} \nonumber \\
&\hspace{-4mm}+(\alpha_1k^2+\gamma_1k+\rho_1)\omega^{10}+\tau^2\omega^{12}\geq0, \;\forall \; \omega\in\mathbb{R}.
\end{align}
Through calculation, we get
\begin{align*}
\left\{
\begin{array}{rll}
\alpha_5k^2-\gamma_5k=&2k_p\beta_3 \overline n_2+{\overline d^2_1}-2k_p\beta_3\overline d_2-\overline n^2_1,\\
\alpha_4k^2+\gamma_4k+\rho_4 = &2\overline n_1\overline n_3+2k_p\beta_3\overline d_4+\overline d^2_2-\overline n^2_2\\
&-2k_pk_v\beta_3-2\overline d_1\overline d_3,\\
\alpha_3k^2+\gamma_3k+\rho_3 =&2\overline n_2k_v+2\overline d_1\overline d_5+\overline d^2_3-\overline n^2_3\\
&-2k_p\beta_3\tau-2\overline d_2 \overline d_4, \\
\alpha_2k^2+\gamma_2k+\rho_2 =& \overline d^2_4+2\overline d_2\tau-k_v^2-2\overline d_3\overline d_5,\\
\alpha_1k^2+\gamma_1k+\rho_1 =& \overline d^2_5-2\overline d_4\tau.
\end{array}
\right.
\end{align*}
This together with (\ref{equ2}) leads to
\begin{align}\label{equ1}
&(2k_p\beta_3\overline n_2+{\overline d^2_1}-2k_p\beta_3\overline d_2-\overline n^2_1)\omega^2 \nonumber \\
&\hspace{-4mm}+(2\overline n_1\overline n_3+2k_p\beta_3\overline d_4+\overline d^2_2-\overline n^2_2-2k_pk_v\beta_3-2\overline d_1\overline d_3)\omega^4 \nonumber \\
&\hspace{-4mm}+(2\overline n_2k_v+2\overline d_1\overline d_5+\overline d^2_3-\overline n^2_3-2k_p\beta_3\tau-2\overline d_2 \overline d_4)\omega^6 \nonumber \\
&\hspace{-4mm}+(\overline d^2_4+2\overline d_2\tau-k_v^2-2\overline d_3\overline d_5)\omega^8 \nonumber \\
&\hspace{-4mm}+(\overline d^2_5-2\overline d_4\tau)\omega^{10}+\tau^2\omega^{12}\geq 0,\;\forall\; \omega\in\mathbb{R}.
\end{align}
Through calculation, we know
% \begin{align*}
%&x_d^2+y_d^2-x_n^2-y_n^2\nonumber\\
%&\hspace{2mm}=\overline d^2_6\omega^{12}+(\overline d^2_5-2\overline d_4\overline d_6)\omega^{10}\nonumber\\
%&\hspace{6mm}+(\overline d^2_4+2\overline d_2\overline d_6-\overline n^2_4-2\overline d_3\overline d_5)\omega^8 \nonumber\\
%&\hspace{6mm}+(2\overline n_2\overline n_4+2\overline d_1\overline d_5+\overline d^2_3-\overline n^2_3-2\overline d_0\overline d_6-2\overline d_2 \overline d_4)\omega^6 \nonumber\\
%&\hspace{6mm}+(2\overline n_1\overline n_3+2\overline d_0\overline d_4+\overline d^2_2-\overline n^2_2-2\overline n_0\overline n_4-2\overline d_1\overline d_3)\omega^4\nonumber\\
%&\hspace{6mm}+(2\overline n_0 \overline n_2+{\overline d^2_1}-2\overline d_0\overline d_2-\overline n^2_1)\omega^2+\overline d^2_0-\overline n^2_0.
%\end{align*}
\begin{align*}
&x_d^2(\omega)+y_d^2(\omega)-x_n^2(\omega)-y_n^2(\omega)\nonumber\\
\hspace{-4mm}=&(2k_p\beta_3\overline n_2+{\overline d^2_1}-2k_p\beta_3\overline d_2-\overline n^2_1)\omega^2 \nonumber \\
\hspace{-4mm}&+(2\overline n_1\overline n_3+2k_p\beta_3\overline d_4+\overline d^2_2-\overline n^2_2-2k_pk_v\beta_3-2\overline d_1\overline d_3)\omega^4 \nonumber \\
\hspace{-4mm}&+(2\overline n_2k_v+2\overline d_1\overline d_5+\overline d^2_3-\overline n^2_3-2k_p\beta_3\tau-2\overline d_2 \overline d_4)\omega^6 \nonumber \\
\hspace{-4mm}&+(\overline d^2_4+2\overline d_2\tau-k_v^2-2\overline d_3\overline d_5)\omega^8 \nonumber \\
\hspace{-4mm}&+(\overline d^2_5-2\overline d_4\tau)\omega^{10}+\tau^2\omega^{12}
\end{align*}
This together with (\ref{equ1}) leads to
\begin{align}\label{equx}
x_d^2(\omega)+y_d^2(\omega)-x_n^2(\omega)-y_n^2(\omega)\geq 0,\;\forall\; \omega\in\mathbb{R}.
\end{align}
By (\ref{equx}), we know
\begin{align}\label{equx1}
\frac{x_n^2(\omega)+y_n^2(\omega)}{x_d^2(\omega)+y_d^2(\omega)}\leq1,\;\forall\; \omega\in\mathbb{R}.
\end{align}
Through calculation, we obtain
\begin{align*}
\left|\frac{x_n(\omega)+y_n(\omega)j}{x_d(\omega)+y_d(\omega)j}\right|=
\frac{\sqrt{x_n^2(\omega)+y_n^2(\omega)}}{\sqrt{x_d^2(\omega)+y_d^2(\omega)}}.
\end{align*}
This together with (\ref{equx1}) leads to
\begin{align}\label{equx2}
\left|\frac{x_n(\omega)+y_n(\omega)j}{x_d(\omega)+y_d(\omega)j}\right| \leq 1,\;\forall\; \omega\in\mathbb{R}.
\end{align}
From (\ref{G_e}) and (\ref{equx2}), we know $|G_{ei}(j\omega)|\leq1$ for any $\omega\in \mathbb{R}$. That is $\|G_{ei}(s)\|_{\infty}\leq1$.\qed

\section{Proof of Theorem 3}\label{Theorem 3}
\textbf{Proof of Theorem 3.} For simplicity of presentation, we denote $b_i=\frac{1}{\tau}+\epsilon_i$. According to (\ref{x_ie}), the models of the velocity difference between adjacent vehicles are given by
\begin{align}\label{v_die}
\left\{
\begin{aligned}
\dot v_{d,i}(t) &= a_{d,i}(t),  \vspace{1ex}\\
\dot a_{d,i}(t) &= q_{1i}(t) +a_i(t)/\tau-u_i(t)/\tau, \quad\quad i=1,2,...,N, \vspace{1ex}\\
\dot q_{1i}(t) &= w_{1i}(t),
\end{aligned}
\right.
\end{align}
where the definitions of $a_{d,i}(t)$ and $v_{d,i}(t)$ are the same as in (\ref{v_di}),
\begin{align}
q_{1i}(t) =& -b_{i-1}a_{i-1}(t)+b_{i-1}u_{i-1}(t)+\epsilon_ia_i(t)\nonumber \\ &-\epsilon_iu_i(t), \label{q_ie} \vspace{1ex}\\
w_{1i}(t) =& b_{i-1}^2a_{i-1}(t)-b_{i-1}^2u_{i-1}(t)+b_{i-1}\dot u_{i-1}(t)\nonumber \\  &-\epsilon_i^2a_i(t)+\epsilon_i^2u_i(t)-\epsilon_i\dot u_i(t).\label{w_ie}
\end{align}
Denote
\begin{align}
X_i(t)=&[p_i(t),v_i(t),a_i(t)]^{T},\;i=0,1,2,...,N\label{x16e}, \vspace{1ex}\\
E_i(t)=&[e_{1,i}(t),e_{2,i}(t),e_{3,i}(t)]^{T},\;i=1,2,...,N\label{e16e}, \vspace{1ex}\\
F_i(t) = &[e_i(t),v_{d,i}(t),a_i(t)]^{T},\;i=1,2,...N, \label{f18e}\vspace{1ex}\\
W(t) = &[F_1^{T}(t),F_2^{T}(t),\cdots,F_N^{T}(t),E_1^{T}(t),E_2^{T}(t), \nonumber \\
&\cdots,E_N^{T}(t)]^{T},\label{We} \vspace{1ex}\\
\Delta(t) = &[\delta_1^{T}(t),0,\cdots,0,\zeta_1^{T}(t),\zeta_2^{T}(t),\zeta_3^{T}(t),0,\nonumber \\
&\cdots,0]^{T},\label{deltae}
\end{align}
where
\begin{align}
e_{1,i}(t)=&z_{1,i}(t)-v_{d,i}(t), \label{e_1ie}\\
e_{2,i}(t)=&z_{2,i}(t)-a_{d,i}(t), \label{e_2ie}\\
e_{3,i}(t)=&z_{3,i}(t)-q_{i\epsilon}(t), \label{e_3ie}\\
\delta_1(t)=& \left[0,a_0(t),b_1k_aa_0(t)\right]^{T}, \label{delta_1e}\\
\zeta_1(t) =& \left[0,0,-b_0^2a_0(t)+b_0^2u_0(t)-b_0\dot u_0(t)+\epsilon_1^2k_aa_0(t)\right.\nonumber \\&\left.-k_v\epsilon_1a_0(t)+k_vk_a\epsilon_1b_1ha_0(t)+\epsilon_1k_a\dot a_0(t)\right]^{T}, \label{zeta_1e}\\
\zeta_2(t) = & \left[0,0,b_1^2k_aa_0(t)+k_a^2\epsilon_2b_1a_0(t)-k_vb_1a_0(t)\right.\nonumber \\&\left.+k_vk_ahb_1^2a_0(t)-b_1k_a\dot a_0(t)\right]^{T}, \label{zeta_2e}\\
\zeta_3(t)=&  \left[0,0,k_a^2b_1b_2a_0(t)\right]^{T}.\label{zeta_3e}
\end{align}
From (\ref{a_di1}), (\ref{u_i1}) and (\ref{e_2ie}), we know
\begin{align}\label{u_i2e}
u_i(t) = &\;k_pe_i(t)+k_v(v_{d,i}(t)-ha_i(t)) \nonumber\\
&+k_a(a_{i-1}(t)+e_{2i}(t)),\;i=1,2,\ldots,N.
\end{align}
This together with (\ref{x_ie}), (\ref{e_i}), (\ref{v_di1}), (\ref{x16e}) and (\ref{e16e}) leads to
\begin{align}\label{dX_ie}
\dot{X_i}(t)=\left\{
\begin{array}{ll}
A_0X_i(t)+B_0u_i(t),&\hspace{-15mm}i=0, \vspace{1ex} \\
A_iX_i(t)+B_iX_{i-1}(t)+C_iE_i(t)+L_{1i}r,\\
&\hspace{-15mm}i=1,\ldots,N,
\end{array}
\right.
\end{align}
where
\begin{equation}
A_0
=\begin{bmatrix}
0 & 1 & 0\\
0 & 0 & 1\\
0 & 0 & \begin{aligned}-b_0\end{aligned}
\end{bmatrix} \nonumber,
B_0
= \begin{bmatrix}
0 \\
0  \\
b_0
\end{bmatrix}\nonumber,
L_{1i}
= \begin{bmatrix}
0 \\
0  \\
\begin{aligned}-k_pb_i\end{aligned}
\end{bmatrix}\nonumber,
\end{equation}
\begin{equation}
A_i
=\begin{bmatrix}
0 & 1 & 0\\
0 & 0 & 1\\
\begin{aligned}-k_pb_i\end{aligned} & \begin{aligned}-k_phb_i-k_vb_i\end{aligned} & \begin{aligned}-b_i-k_vhb_i\end{aligned}
\end{bmatrix},  \nonumber
\end{equation}
\begin{equation}
B_i
= \begin{bmatrix}
0 & 0 & 0\\
0 & 0 & 0 \\
\begin{aligned}k_pb_i\end{aligned} & \begin{aligned}k_vb_i\end{aligned} & \begin{aligned}k_ab_i\end{aligned}
\end{bmatrix} \nonumber,
C_i
=\begin{bmatrix}
0 & 0 & 0 \\
0 & 0 & 0 \\
0 & \begin{aligned}k_ab_i\end{aligned} & 0
\end{bmatrix} \nonumber.
\end{equation}

From (\ref{e_i}), (\ref{x16e}) and (\ref{f18e}), we get
\begin{eqnarray}\label{F_ie}
F_i(t)=PX_{i-1}(t) - QX_i(t)-L_{2i}r,\;i=1,2,\ldots,N,
\end{eqnarray}
where
\begin{equation}
P = {
	\left[ \begin{array}{ccc}
	1 & 0 & 0\\
	0 & 1 & 0\\
	0 & 0 & 0
	\end{array}
	\right ]} \nonumber,
Q = {
	\left[ \begin{array}{ccc}
	1 & h & 0\\
	0 & 1 & 0\\
	0 & 0 & -1
	\end{array}
	\right ]} \nonumber,
L_{2i} = {
	\left[ \begin{array}{ccc}
	1 \\
	0 \\
	0
	\end{array}
	\right ]}. \nonumber
\end{equation}
From (\ref{delta_1e}), (\ref{dX_ie}) and (\ref{F_ie}), we obtain
\begin{align}\label{dF_ie}
\dot{F_i}(t)=\left\{
\begin{array}{ll}
\mathcal{A}_{1i} F_i(t)+\mathcal{G}_iE_i(t)+\delta_i(t),&\hspace{-1.0mm}i = 1, \vspace{1ex} \\
\mathcal{A}_{1i} F_i(t)+\mathcal{B}_{1i}F_{i-1}(t)+\mathcal{G}_iE_i(t),&\hspace{-1.0mm}i=2,\ldots,N,
\end{array}
\right.
\end{align}
where
\begin{equation}
\mathcal{A}_{1i}
=\begin{bmatrix}
%\begin{smallmatrix}
0 & 1 & -h\\
0 & 0 & -1\\
\begin{aligned}k_pb_i\end{aligned} & \begin{aligned}k_vb_i\end{aligned} &\begin{aligned}(-1-k_vh)b_i\end{aligned}
%\end{smallmatrix}
\end{bmatrix} \nonumber,
\end{equation}
\begin{equation}
\mathcal{B}_{1i}
= \begin{bmatrix}
0 & 0 & 0\\
0 & 0 & 1 \\
0 & 0 & \begin{aligned}b_ik_a\end{aligned}
\end{bmatrix} \nonumber,
\mathcal{G}_i
= \begin{bmatrix}
0 & 0 & 0\\
0 & 0 & 0 \\
0 & \begin{aligned}b_ik_a\end{aligned} & 0
\end{bmatrix}.\nonumber
\end{equation}
By (\ref{e_i}), (\ref{v_di1}), (\ref{a_di1}), (\ref{z_i}), (\ref{x_ie}), (\ref{q_ie})-(\ref{f18e}), (\ref{e_1ie})-(\ref{u_i2e}) and (\ref{dF_ie}), we have
\begin{align}\label{dE_ie}
\dot{E_i}(t)=\left\{
\begin{array}{ll}
\mathcal{C}_{i} F_i(t)+\mathcal{H}_{1i}E_i(t)+\zeta_i(t) , &\hspace{-8mm}i = 1, \vspace{1ex}\\
\mathcal{C}_{i} F_i(t)+\mathcal{D}_{1i} F_{i-1}(t)+\mathcal{H}_{1i}E_{i}(t)\\
+\mathcal{I}_iE_{i-1}(t)+\zeta_i(t) , &\hspace{-8mm}i = 2, \vspace{1ex}\\
\mathcal{C}_{i} F_i(t)+\mathcal{D}_{1i} F_{i-1}(t)+\mathcal{E}_{1i} F_{i-2}(t)\\
+\mathcal{H}_{1i}E_i(t)+\mathcal{I}_iE_{i-1}(t)+\mathcal{J}_iE_{i-2}(t)\\
+\zeta_i(t) , &\hspace{-8mm}i = 3,\vspace{1ex}\\
\mathcal{C}_{i} F_i(t)+\mathcal{D}_{1i} F_{i-1}(t)+\mathcal{E}_{1i} F_{i-2}(t)\\
+\mathcal{F}_iF_{i-3}(t)+\mathcal{H}_{1i}E_i(t)+\mathcal{I}_iE_{i-1}(t)\\
+\mathcal{J}_iE_{i-2}(t),  &\hspace{-8mm}i=4,\ldots,N,
\end{array}
\right.
\end{align}
where
\begin{equation}
\mathcal{H}_{1i}
= \begin{bmatrix}
-\beta_1 & 1 & 0 \\
-\beta_2 & 0 & 1 \\
-\beta_3-k_a\beta_2\epsilon_i\ & \begin{aligned}-k_a\epsilon_i^2-k_vk_ahb_i\epsilon_i\end{aligned} & k_a\epsilon_i
\end{bmatrix}, \nonumber
\end{equation}
\begin{equation}
\mathcal{F}_i
=\begin{bmatrix}
0 & 0 & 0\\
0 & 0 & 0\\
0 & 0 & \begin{aligned}-k_a^2b_{i-1}b_{i-2}\end{aligned}
\end{bmatrix}, \nonumber\\
\mathcal{J}_i
= \begin{bmatrix}
0 & 0 & 0 \\
0 & 0 & 0 \\
0 & \begin{aligned}-k_a^2b_{i-1}b_{i-2}\end{aligned} & 0
\end{bmatrix}, \nonumber
\end{equation}
\begin{equation}
\mathcal{C}_{i}
=\begin{bmatrix}
0 & 0 & 0\\
0 & 0 & 0\\
\begin{aligned}&-k_pk_vhb_i\epsilon_i\\&-k_p\epsilon_i^2\end{aligned} & \begin{aligned}&k_p\epsilon_i-k_v\epsilon_i^2\\&-k_v^2hb_i\epsilon_i\end{aligned} & \begin{aligned}&\epsilon_i^2+k_vh\epsilon_i^2-k_v\epsilon_i\\
&-k_ph\epsilon_i+k_vhb_i\epsilon_i\\&+k_v^2h^2b_i\epsilon_i\end{aligned}
\end{bmatrix}, \nonumber\\
\end{equation}
\begin{equation}
\mathcal{D}_{1i}
=\begin{bmatrix}
0 & 0 & 0\\
0 & 0 & 0\\
\begin{aligned}&k_pk_vhb_{i-1}^2\\&+k_pk_a\epsilon_ib_{i-1}\\&+k_pb_{i-1}^2\end{aligned} & \begin{aligned}&k_vk_a\epsilon_ib_{i-1}\\&+k_v^2hb_{i-1}^2\\&-k_pb_{i-1}\\&+k_vb_{i-1}^2\end{aligned} & \begin{aligned}&k_phb_{i-1}+k_vb_{i-1}\\&-b_{i-1}^2-k_a\epsilon_ib_{i-1}\\&-k_v^2h^2b_{i-1}^2-k_a\epsilon_i^2\\&+k_v\epsilon_i-k_vk_ah\epsilon_ib_i\\&-k_vk_ah\epsilon_ib_{i-1}\\&-2k_vhb_{i-1}^2\end{aligned}
\end{bmatrix}, \nonumber
\end{equation}
\begin{equation}
\mathcal{E}_{1i}
= \begin{bmatrix}
0 & 0 & 0\\
0 & 0 & 0 \\
\begin{aligned}-k_pk_ab_{i-1}b_{i-2}\end{aligned} & \begin{aligned}-k_vk_ab_{i-1}b_{i-2}\end{aligned} & \begin{aligned}&k_vk_ahb_{i-1}b_{i-2}\\&+k_vk_ahb_{i-1}^2\\&+k_ab_{i-1}b_{i-2}\\&+k_a^2\epsilon_ib_{i-1}\\&+k_ab_{i-1}^2\\&-k_vb_{i-1}\end{aligned}
\end{bmatrix}, \nonumber
\end{equation}
\begin{equation}
\mathcal{I}_i
= \begin{bmatrix}
0 & 0 & 0\\
0 & 0 & 0 \\
\begin{aligned}k_a\beta_2b_{i-1}\end{aligned} & \begin{aligned}&k_ab_{i-1}^2+k_vk_ahb_{i-1}^2\\&+k_a^2\epsilon_ib_{i-1}\end{aligned} & \begin{aligned}-k_ab_{i-1}\end{aligned}
\end{bmatrix}. \nonumber
\end{equation}
From (\ref{We}), (\ref{deltae}), (\ref{dF_ie}) and (\ref{dE_ie}), we know
\begin{align}\label{closed-loop-3}
\dot{W}(t) = (\Psi+ \hat \Psi)W(t)+\Delta(t),
\end{align}
where
$\hat \Psi
=\begin{bmatrix}
\hat \Psi_{11} & \hat \Psi_{12}  \\
\hat \Psi_{21} & \hat \Psi_{22}
\end{bmatrix}$ and
\begin{equation}
\hat \Psi_{11}
=\begin{bmatrix}
%\begin{smallmatrix}
\mathcal {A}_{21} & \quad & \quad & \quad   \\
\mathcal {B}_{22} & \mathcal {A}_{22} & \quad & \quad  \\
\quad & \ddots & \ddots & \quad  \\
\quad & \quad  & \mathcal {B}_{2N} & \mathcal {A}_{2N} \\
%\end{smallmatrix}
\end{bmatrix},\nonumber
\hat \Psi_{12}
=\begin{bmatrix}
%\begin{smallmatrix}
\mathcal {G}_1 & \quad & \quad  \\
\quad & \ddots & \quad  \\
\quad & \quad  &  \mathcal {G}_N \\
%\end{smallmatrix}
\end{bmatrix},\nonumber
\end{equation}
\begin{equation}
\hat \Psi_{21}
=\begin{bmatrix}
%\begin{smallmatrix}
\mathcal {C}_{1}& \quad & \quad & \quad & \quad & \quad \\
\mathcal {D}_{22} & \mathcal {C}_{2} & \quad & \quad & \quad & \quad \\
\mathcal {E}_{23} & \mathcal {D}_{23} & \mathcal {C}_{3} & \quad & \quad & \quad \\
\mathcal {F}_4 & \mathcal {E}_{24} & \mathcal {D}_{24} & \mathcal {C}_{4} & \quad & \quad \\
\quad & \ddots & \ddots & \ddots & \ddots & \quad \\
\quad  & \quad & \mathcal {F}_N & \mathcal {E}_{2N} & \mathcal {D}_{2N} & \mathcal {C}_{N}
%\end{smallmatrix}
\end{bmatrix},\nonumber
\end{equation}
\begin{equation}
\hat \Psi_{22}
=\begin{bmatrix}
%\begin{smallmatrix}
\mathcal {H}_{21} & \quad & \quad & \quad & \quad \\
\mathcal {I}_2 & \mathcal {H}_{22}& \quad & \quad & \quad \\
\mathcal {J}_3 & \mathcal {I}_3 & \mathcal {H}_{23} & \quad & \quad \\
\quad & \ddots & \ddots & \ddots & \quad \\
\quad & \quad & \mathcal {J}_N  & \mathcal {I}_N & \mathcal {H}_{2N}
%\end{smallmatrix}
\end{bmatrix}, \nonumber
\end{equation}
\begin{align*}
&\mathcal {A}_{2i}=\mathcal {A}_{1i}-\mathcal {A}, \mathcal {B}_{2i}=\mathcal {B}_{1i}-\mathcal {B},\mathcal {D}_{2i}=\mathcal {D}_{1i}-\mathcal {D},\\
 & \mathcal {E}_{2i}=\mathcal {E}_{1i}-\mathcal {E}, \mathcal {H}_{2i}=\mathcal {H}_{1i}-\mathcal {H}.
\end{align*}
Firstly, we analyze the stability of $\Psi$. The eigenvalues of $\Psi$ are only related to $\mathcal{A}$ and $\mathcal{H}$. Calculating the characteristic polynomial of $\mathcal{A}$, we obtain
\begin{align}\label{A_de}
\left|sI- \mathcal{A} \right|= s^3+\left(\frac{1+k_vh}{\tau}\right)s^2+\left(\frac{k_v+k_ph}{\tau}\right)s+\frac{k_p}{\tau}.
\end{align}
The Rouse table corresponding to (\ref{A_de}) is given by
\begin{table}[H]
	\centering
	\begin{tabular}{c c c }
		$s^3$ & 1 & $\begin{aligned}\frac{k_v+k_ph}{\tau}\end{aligned}$ \\
		$s^2$ & $\begin{aligned}\frac{1+k_vh}{\tau}\end{aligned}$ & $\begin{aligned}\frac{k_p}{\tau}\end{aligned}$  \\
		$s^1$ & $\begin{aligned}\frac{hk_v^2+(1+h^2k_p)k_v+(h-\tau)k_p}{\tau+\tau k_vh}\end{aligned}$ & 0  \\
		$s^0$ & $\begin{aligned}\frac{k_p}{\tau}\end{aligned}$ &  \\
	\end{tabular}
\end{table}
By $k_p>0$ and (\ref{h3e}), we know that the elements of the first column of the Rouse table corresponding to (\ref{A_de}) are all greater than zero. From Rouse criterion, $\mathcal{A}$ is stable.
Calculating the characteristic polynomial of $\mathcal{H}$, we get
\begin{align}\label{Se}
\left|sI- \mathcal{H}\right| = s^3+\beta_1s^2+\beta_2s+\beta_3.
\end{align}
The Rouse table corresponding to (\ref{Se}) is given by
\begin{table}[H]
	\centering
	\begin{tabular}{c c c }
		$s^3$ & 1 & $\beta_2$ \\
		$s^2$ & $\beta_1$ & $\beta_3$  \\
		$s^1$ & $\begin{aligned}\frac{\beta_1 \beta_2-\beta_3}{\beta_1}\end{aligned}$ & 0  \\
		$s^0$ & $\beta_3$ &
	\end{tabular}
\end{table}
By $\beta_1>0$, $\beta_3>0$, $\beta_1\beta_2-\beta_3>0$, we know that the elements of the first column of the Rouse table corresponding to (\ref{Se}) are all greater than zero. From Rouse criterion, $\mathcal{H}$ is stable. Then $\Psi$ is stable.

From the definition of $\|\hat{\Psi}\|$ and Lemma 2, we know
\begin{align}\label{hatpsie}
\|\hat{\Psi}\|\leq\left\{
\begin{array}{lc}
\begin{aligned}&\left(1/\tau+Y_1\overline \epsilon^2+Y_2\overline \epsilon\right)k_a+Z_1\overline\epsilon^2+Z_2\overline\epsilon,\\&\hspace{54mm}{\rm if} \ N=1,\end{aligned} \vspace{1ex}\\
\begin{aligned}&A_1k_a^2+\left(\Theta_1+Y_3\overline \epsilon^2+Y_4\overline \epsilon\right)k_a+Z_3\overline \epsilon^2+Z_4\overline \epsilon,\\&\hspace{54mm}{\rm if} \ N=2,\end{aligned} \vspace{1ex}\\
\begin{aligned}
&A_2k_a^2+(\Theta+Y_5\overline \epsilon^2+Y_6\overline \epsilon)k_a+Z_5\overline \epsilon^2+Z_6\overline \epsilon,\\ &\hspace{54mm}{\rm if} \ N\geq3.
\end{aligned}
\end{array}
\right.
\end{align}
From (\ref{oe1}), (\ref{k_ae}) and (\ref{hatpsie}), we get $\|\hat{\Psi}\|< r_c(\Psi)$. It is known from the definition of $\Delta(t)$ and Assumption 1 that $\lim \limits_ {t\to \infty}\Delta(t)=0$. By Lemma 1 and (\ref{closed-loop-3}), we know $W(t)$ converges to zero exponentially. Then $v_{i}(t)-v_0(t)$ and $e_i(t)$ both converge to zero exponentially, $i=1,2,...,N$.  \qed

\section{Proof of Theorem 4}\label{Theorem 4}
\textbf{Proof of Theorem 4.}
For simplicity of presentation, we denote $b_i=\frac{1}{\tau}+\epsilon_i$.
By (\ref{e_i}) and (\ref{v_di1}), we get
\begin{align}\label{a_ise}
a_i(t) = \frac{v_{d,i}(t)- \dot e_i(t)}{h}.
\end{align}
Taking the Laplace transform of (\ref{a_ise}), we have
\begin{align}\label{La_ise}
\mathscr{A}_i(s) = \frac{\mathscr{V}_{d,i}(s)- s\mathscr{E}_i(s)}{h},
\end{align}
where $\mathscr{A}_i(s)$ and $\mathscr{V}_{d,i}(s)$ are the Laplace transform of $a_i(t)$, $v_{d,i}(t)$, respectively.
From (\ref{x_ie}), we know
\begin{align}\label{u_i3e}
u_i(t)=\frac{\dot a_i(t)}{b_i} + a_i(t).
\end{align}
This together with (\ref{a_ise}) leads to
\begin{align}\label{u_ise}
u_i(t)=\frac{\dot v_{d,i}(t)- \ddot e_i(t)}{b_ih}+\frac{v_{d,i}(t)- \dot e_i(t)}{h}.
\end{align}
Taking the Laplace transform of (\ref{u_ise}), we get
\begin{align}\label{Lu_ise}
\mathscr{U}_i(s)=\frac{s\mathscr{V}_{d,i}(s)- s^2\mathscr{E}_i(s)}{b_ih}+\frac{\mathscr{V}_{d,i}(s)- s\mathscr{E}_i(s)}{h},
\end{align}
where $\mathscr{U}_i(s)$ is the Laplace transform of $u_i(t)$.
Taking the Laplace transform of (\ref{z_i}), we have
\begin{align}\label{z_ise}
\left\{
\begin{array}{lll} %Éè¶šÁÐÕóµÄžñÊœ£º{lll} ÊÇÈýžöL£¬±íÊŸÈýÁÐµÄ¶ÔÆë·œÊœÎªLeft ¶ÔÆë
s\mathscr{Z}_{1,i}(s) = &\mathscr{Z}_{2,i}(s) +\beta_{1}(\mathscr{V}_{d,i}(s)-\mathscr{Z}_{1,i}(s)), \vspace{1ex} \\
s\mathscr{Z}_{2,i}(s) = &\mathscr{Z}_{3,i}(s) +\beta_{2}(\mathscr{V}_{d,i}(s)-\mathscr{Z}_{1,i}(s))+\mathscr{A}_i(s)/\tau\\
&-\mathscr{U}_i(s)/\tau, \vspace{1ex}  \\
s\mathscr{Z}_{3,i}(s) = &\beta_{3}(\mathscr{V}_{d,i}(s)-\mathscr{Z}_{1,i}(s)),\vspace{1ex}
\end{array} %·œ³ÌÁÐÕóµÄœáÊø
\right. %·œ³Ì×éµÄÓÒ±ßÎÞ·ûºÅ£¬ÀûÓÃ¡°.¡°ÀŽ±êÊŸ
\end{align}
%\begin{center}
%$i = 1,2,\ldots,N$,
%\end{center}
where $\mathscr{Z}_{1,i}(s)$, $\mathscr{Z}_{2,i}(s)$ and $\mathscr{Z}_{3,i}(s)$ are the Laplace transform of $z_{1,i}(t)$, $z_{2,i}(t)$ and $z_{3,i}(t)$, respectively.
Substituting (\ref{Lu_ise}) into (\ref{z_ise}), we obtain
\begin{align}\label{z_2is}
\mathscr{Z}_{2,i}(s) =&\; ((-s^3+(\tau b_i h\beta_2 -\beta_1)s^2+\tau b_ih\beta_3s)\mathscr{V}_{d,i}(s)\nonumber \\
&+(s^4+\beta_1s^3)\mathscr{E}_i(s))/(\tau b_ih(s^3+\beta_1s^2\nonumber  \\
&+\beta_2s+\beta_3))
\end{align}
By (\ref{u_i1}), (\ref{a_ise}) and (\ref{u_i3e}), we get
\begin{align}\label{22e}
\frac{\dot a_i(t)}{b_i}+a_i(t) = &\;k_{p}e_i(t)+k_{v}\dot e_i(t) \nonumber \\
&+k_{a}(z_{2,i}(t)+a_i(t)).
\end{align}
Taking the Laplace transform of (\ref{22e}), we have
\begin{align}\label{2e}
\frac{s\mathscr{A}_i(s)}{b_i}+\mathscr{A}_i(s)=&\;k_p\mathscr{E}_i(s)+k_v s\mathscr{E}_i(s)+k_a\mathscr{A}_i(s)\nonumber\\
&+k_a\mathscr{Z}_{2,i}(s).
\end{align}
Denote $H_i(s)=\frac{\mathscr{V}_{d,i}(s)}{\mathscr{E}_i(s)}$. By (\ref{La_ise}), (\ref{z_2is}) and (\ref{2e}), we get
\begin{align}\label{HHe}
H_i(s)=\frac{ s^5+ n_{4i}s^4+ n_{3i}s^3+ n_{2i}s^2+ n_{1i}s+ n_{0i}}{ s^4+d_{3i}s^3+d_{2i}s^2+d_{1i}s+d_{0i}},
\end{align}
where
\begin{align*}
    n_{0i} = & b_ik_ph\beta_3,  \\
	n_{1i} = & b_ik_ph\beta_2+( b_i- b_ik_a+ b_ik_vh)\beta_3, \\
	n_{2i} = & b_ik_ph\beta_1+( b_i- b_ik_a+ b_ik_vh)\beta_2+\beta_3,\nonumber \\
	n_{3i} = &\ b_ik_ph+(k_a/\tau+ b_i- b_ik_a+ b_ik_vh)\beta_1+\beta_2, \\
	n_{4i} = &k_a/\tau+ b_i- b_ik_a+ b_ik_vh+ \beta_1 ,\\
	d_{0i} = &( b_i- b_ik_a)\beta_3,\\
	d_{1i} = &( b_i- b_ik_a)\beta_2+(1- b_ik_ah)\beta_3,\nonumber \\
	d_{2i}= &(k_a/\tau- b_ik_a+ b_i)\beta_1+(1- b_ik_ah)\beta_2,  \\
	d_{3i} =& k_a/\tau- b_ik_a+ b_i+ \beta_1.
\end{align*}
By (\ref{e_i}) and (\ref{v_di1}), we get
\begin{align}\label{dee}
\dot e_{i-1}(t)- \dot e_i(t)=v_{d,i-1}(t)-v_{d,i}(t)-h\dot v_{d,i}(t).
\end{align}
Denote $G_{ei}(s)=\frac{\mathscr{E}_i(s)}{\mathscr{E}_{i-1}(s)}$. Taking the Laplace transform of (\ref{dee}), we have
\begin{align*}%\label{Lde}
s\mathscr{E}_{i-1}(t)- s\mathscr{E}_i(t)=\mathscr{V}_{d,i-1}(t)-\mathscr{V}_{d,i}(t)-hs\mathscr{V}_{d,i}(t).
\end{align*}
This together with $\mathscr{V}_{d,i}(s)=H_i(s)\mathscr{E}_i(s)$ leads to
\begin{align*}%\label{G_es1}
G_{ei}(s) =\frac{s-H_i(s)}{s-(hs+1)H_i(s)}.
\end{align*}
This together with (\ref{HHe}) leads to
%\begin{align}\label{G_es}
%&G_e(s) \nonumber\\
% &=\frac{ \overline n_4s^4+\overline n_3s^3+ \overline n_2s^2+ \overline n_1s+ \overline n_0}{\overline d_6s^6+\overline d_5s^5+\overline d_4s^4+\overline d_3s^3+\overline d_2s^2+\overline d_1s+\overline d_0},
%\end{align}
\begin{align}\label{G_ese}
G_{ei}(s)= &\;( b_ik_vs^4+\overline n_{3i}s^3+ \overline n_{2i}s^2+ \overline n_{1i}s+ b_ik_p\beta_3) \nonumber  \\
&/( s^6 \nonumber+\overline d_{5i}s^5+\overline d_{4i}s^4+\overline d_{3i}s^3+\overline d_{2i}s^2+\overline d_{1i}s \nonumber \\
&+b_ik_p\beta_3),
\end{align}
where
\begin{align*}
	\overline n_{1i} = & b_ik_p\beta_2+ b_ik_v\beta_3,\\
	\overline n_{2i} = & b_ik_p\beta_1+ b_ik_v\beta_2+ b_ik_a\beta_3, \\
	\overline n_{3i} = & b_ik_v\beta_1+ b_ik_a\beta_2+ b_ik_p,  \\
    \overline d_{1i} = & b_ik_p\beta_2+( b_ik_ph+ b_ik_v)\beta_3,  \\
	\overline d_{2i} = & b_ik_p\beta_1+( b_ik_ph+ b_ik_v)\beta_2+(b_i+ b_ik_vh)\beta_3,\\
    \overline d_{3i} = &(b_ik_ph+ b_ik_v)\beta_1+(b_i+b_ik_vh)\beta_2\\
                      &+\beta_3+b_ik_p,\\
	\overline d_{4i} = &(k_a/\tau- b_ik_a+ b_i+ b_ik_vh)\beta_1+  \beta_2\\
                       &+ b_ik_ph+ b_i k_v\\
	\overline d_{5i} = & \beta_1+k_a/\tau- b_ik_a+ b_i+ b_ik_vh.
\end{align*}
Substituting $s=j\omega$ into (\ref{G_ese}), we get
\begin{align}\label{G_ee}
G_{ei}(j\omega)=\frac{x_{ni}(\omega)+y_{ni}(\omega)j}{x_{di}(\omega)+y_{di}(\omega)j},
\end{align}
where $x_{ni}(\omega)=b_ik_p\beta_3- \overline n_{2i}\omega^2+ b_ik_v\omega^4$, $y_{ni}(\omega) =\overline n_{1i}\omega- \overline n_{3i}\omega^3 $, $x_{di}(\omega)= b_ik_p\beta_3- \overline d_{2i}\omega^2+\overline d_{4i}\omega^4-\omega^6$, $y_{di}(\omega)=\overline d_{1i}\omega-\overline d_{3i}\omega^3+\overline d_{5i}\omega^5$.

Through calculation, we get
\begin{align}\label{equx0e}
\left\{
\begin{array}{rll}
\alpha_{5i}k^2-\gamma_{5i}k=&2b_ik_p\beta_3 \overline n_{2i}+{\overline d^2_{1i}}-2 b_ik_p\beta_3\overline d_{2i}\\
&-\overline n^2_{1i},\\
\alpha_{4i}k^2+\gamma_{4i}k+\rho_{4i} = &2\overline n_{1i}\overline n_{3i}+2 b_ik_p\beta_3\overline d_{4i}+\overline d^2_{2i}\\
&-\overline n^2_{2i}-2b_i^2k_pk_v\beta_3-2\overline d_{1i}\overline d_{3i},\\
\alpha_{3i}k^2+\gamma_{3i}k+\rho_{3i} =&2\overline n_{2i}b_ik_v+2\overline d_{1i}\overline d_{5i}+\overline d^2_{3i}-\overline n^2_{3i}\\
&-2b_ik_p\beta_3-2\overline d_{2i} \overline d_{4i}, \\
\alpha_{2i}k^2+\gamma_{2i}k+\rho_{2i} =& \overline d^2_{4i}+2\overline d_{2i}-b_i^2k_v^2-2\overline d_{3i}\overline d_{5i},\\
\alpha_{1i}k^2+\gamma_{1i}k+\rho_{1i} =& \overline d^2_{5i}-2\overline d_{4i}.
\end{array}
\right.
\end{align}
where $\rho_{1i} = 3\omega_o^2+b_i^2$, $\rho_{2i} = 3\omega_o^4+3b_i^2\omega_o^2$, $\rho_{3i} = \omega_o^6+3b_i^2\omega_o^4$, $\rho_{4i} =b_i^2\omega_o^6$ and
\begin{align*}
	\lambda_{1i} = &3b_i\mu_vh(b_ih\mu_v-2\mu_a/\tau)+3b_i^2\mu_a(2\mu_vh-3\mu_a),\\
	\lambda_{2i}= &16b^2_ih\mu_a\mu_p-(16b_i\mu_a\mu_v+16b_ih\mu_a\mu_p)/\tau,\\
	\lambda_{3i}= &3b^2_ih^2\mu_p^2+6b^2_i\mu_a\mu_p-12b_i\mu_p\mu_a/\tau\\
    \alpha_{1i} = &(\mu_a/\tau-b_i\mu_a+b_ih\mu_v)^2,\\
	\gamma_{1i} = &2b_i\mu_a/\tau+2b_i^2h\mu_v-2b_i^2\mu_a-2b_ih\mu_p-2b_i\mu_v,\\
\alpha_{2i} =&((12b_ih\mu_a\mu_v-18b_i\mu_a^2)/\tau+9\mu_a^2/\tau^2+9b^2_i\mu_a^2\\
              &+3b^2_ih^2\mu_v^2-12b^2_ih\mu_a\mu_v)\omega_o^2+b^2_ih^2\mu_p^2\\
              &+2b^2_i\mu_a\mu_p-2b_i\mu_a\mu_p/\tau\\
   \gamma_{2i} =&(16\mu_a/\tau-16b_i\mu_a)\omega_o^3+(12b_i\mu_a/\tau+6b_i^2h\mu_v\\
              &-12b_i^2\mu_a-6b_ih\mu_p-6b_i\mu_v)\omega_o^2-2b_i^2\mu_p\\
	\alpha_{3i} = &\lambda_1\omega_o^4+\lambda_2\omega_o^3+\lambda_3\omega_o^2,\\
\gamma_{3i} = &(6b_i^2\mu_a-6b_i\mu_v-6b_ih\mu_p-6b_i\mu_a/\tau+6b_i^2h\mu_v)\omega_o^4\\
                  &-6b_i^2\mu_p\omega_o^2,\\
	\alpha_{4i} = &b^2_i(h^2\mu^2_v-\mu^2_a)\omega_o^6+(3b_i^2h^2\mu_p^2+6b_i^2\mu_p\mu_a+\\
               &6b_i\mu_p\mu_a/\tau)\omega_o^4,\\
	\gamma_{4i} =& (2b_i^2h\mu_v-2b_ih\mu_p-2b_i\mu_v)\omega_o^6-6b_i^2\mu_p\omega_o^4,  \\
    \alpha_{5i} =& (b_i^2h^2\mu_p^2+2b^2_i\mu_a\mu_p)\omega_o^6,\gamma_{5i} =2b^2_i\mu_p\omega_o^6.
    \end{align*}
By $\mu_p>0$ and $\mu_a>0$, we know $\overline \alpha_{5}>0$ and $\underline \gamma_{5}>0$. From (\ref{ke}), $\overline \alpha_{5}>\alpha_{5i}$ and $\gamma_{5i}>\underline \gamma_5$, we obtain $k\geq\gamma_{5i}/\alpha_{5i}$. This together with $\alpha_{5i}>0$ and $\gamma_{5i}>0$ leads to
\begin{align}\label{k5e}
\alpha_{5i}k^2-\gamma_{5i}k\geq0.
\end{align}
From (\ref{mu_ve}) and $\mu_a>0$, we know $\mu_v>\overline b\mu_a/(\underline bh)$. By (\ref{omegae}), we know $\omega_o>((3\underline b^2h^2\mu_p^2+6\underline b^2\mu_p\mu_a+6\underline b\mu_p\mu_a/\tau)/(\underline b^2h^2\mu^2_v-\overline b^2\mu^2_a))^{1/2}$. This together with $\mu_v>\overline b\mu_a/(\underline b h)$ leads to $\underline \alpha_4>0$. This together with $\underline \alpha_4<\alpha_{4i}<\overline \alpha_4$, we know $\alpha_{4i}>0$. From (\ref{ke}) and $\overline \rho_4>\rho_{4i}>0$, we know $k\geq\max\{0,(-\gamma_{4i}+\sqrt{|\gamma_{4i}^2-4\alpha_{4i}\rho_{4i}|})/(2\alpha_{4i})\}$. This together with $\alpha_{4i}>0$ and $\rho_{4i}>0$ leads to
\begin{align}\label{k4e}
\alpha_{4i}k^2+\gamma_{4i}k+\rho_{4i}\geq0.
\end{align}
By (\ref{mu_ve}), we know $\mu_v>\max\{3\mu_a\overline b^2/(2h\underline b^2),2\mu_a\overline b/(\tau h\underline b^2)\}$. This together with $\mu_a>0$ leads to  $\underline \lambda_1>0$. By $\mu_p>4\mu_a\overline b/(\tau\underline b^2h^2)$, we know $3\mu_p(\underline b^2h^2\mu_p-4\mu_a\overline b/\tau)>0$. This together with $\mu_a>0$ leads to $\underline \lambda_3>0$. By (\ref{omegae}), we know $\omega_o>\max\{0,(-\underline \lambda_{2}+\sqrt{|\underline \lambda_{2}^2-4\underline \lambda_{1}\underline \lambda_{3}|})/(2\underline \lambda_{1})\}$. This together with $\underline \lambda_{1}>0$ and $\underline \lambda_{3}>0$ leads to $\underline \alpha_{3}>0$. This together with $\underline \alpha_3<\alpha_{3i}<\overline \alpha_3$, we know $\alpha_{3i}>0$. From (\ref{ke}) and $\overline \rho_3>\rho_{3i}>0$, we know $k\geq\max\{0,(-\gamma_{3i}+\sqrt{|\gamma_{3i}^2-4\alpha_{3i}\rho_{3i}|})/(2\alpha_{3i})\}$. This together with $\alpha_{3i}>0$ and $\rho_{3i}>0$ leads to
\begin{align}\label{k3e}
\alpha_{3i}k^2+\gamma_{3i}k+\rho_{3i}\geq0.
\end{align}
By (\ref{mu_ve}), we know $\mu_v>4\mu_a\overline b^2/(h\underline b^2)$. This leads to $3\underline b^2h^2\mu_v^2-12\overline b^2h\mu_a\mu_v>0$. By (\ref{mu_ve}), we know $\mu_v>3\mu_a\overline b/(2h\underline b)$. This leads to $(12\underline bh\mu_a\mu_v-18\overline b\mu_a^2)/\tau>0$.  By $\mu_p> 2\mu_a\overline b/(\underline b^2h^2)$, we know $\mu_p(\underline b^2h^2\mu_p-2\overline b\mu_a/\tau)>0$. The above together with $\mu_p>0$,$\mu_a>0$ lead to $\underline \alpha_2>0$. This together with $\underline \alpha_2<\alpha_{2i}<\overline \alpha_2$, we know $\alpha_{2i}>0$. From (\ref{ke}) and $\overline \rho_2>\rho_{2i}>0$, we obtain $k\geq\max\{0,(-\gamma_{2i}+\sqrt{|\gamma_{2i}^2-4\alpha_{2i}\rho_{2i}|})/(2\alpha_{2i})\}$. This together with $\alpha_{2i}>0$ and $\rho_{2i}>0$ leads to
\begin{align}\label{k2e}
\alpha_{2i}k^2+\gamma_{2i}k+\rho_{2i}\geq0.
\end{align}
By (\ref{mu_ve}), we know $\mu_v>\mu_a\overline b/(h \underline b)$. This together with $\mu_a>0$, we know $\mu_a/\tau-\overline b\mu_a+\underline bh\mu_v>0$ and $\underline \alpha_1<\alpha_{1i}<\overline \alpha_1$. From (\ref{ke}) and $\overline \rho_1>\rho_{1i}>0$, we obtain $k\geq\max\{0,(-\gamma_{1i}+\sqrt{|\gamma_{1i}^2-4\alpha_{1i}\rho_{1i}|})/(2\alpha_{1i})\}$. This together with $\alpha_{1i}>0$ and $\rho_{1i}>0$ leads to
\begin{align}\label{k1e}
\alpha_{1i}k^2+\gamma_{1i}k+\rho_{1i}\geq0.
\end{align}
By (\ref{k5e})-(\ref{k1e}), we know
\begin{align}\label{equ2e}
&(\alpha_{5i}k^2-\gamma_{5i}k)\omega^{2}+(\alpha_{4i}k^2+\gamma_{4i}k+\rho_{4i})\omega^{4} \nonumber \\
&\hspace{-4mm}+(\alpha_{3i}k^2+\gamma_{3i}k+\rho_{3i})\omega^{6}+(\alpha_{2i}k^2+\gamma_{2i}k+\rho_{2i})\omega^{8} \nonumber \\
&\hspace{-4mm}+(\alpha_{1i}k^2+\gamma_{1i}k+\rho_{1i})\omega^{10}+\omega^{12}\geq0, \;\forall \; \omega\in\mathbb{R}.
\end{align}
By (\ref{equx0e}) and (\ref{equ2e}), we get
\begin{align}\label{equ1e}
&\hspace{4mm}(2b_ik_p\beta_3 \overline n_{2i}+{\overline d^2_{1i}}-2b_ik_p\beta_3\overline d_{2i}-\overline n^2_{1i})\omega^2 \nonumber \\
&+(2\overline n_{1i}\overline n_{3i}+2b_ik_p\beta_3\overline d_{4i}+\overline d^2_{2i}-\overline n^2_{2i}-2b_i^2k_pk_v\beta_3\nonumber \\
&-2\overline d_{1i}\overline d_{3i})\omega^4+(2\overline n_{2i}b_ik_v+2\overline d_{1i}\overline d_{5i}+\overline d^2_{3i}-\overline n^2_{3i} \nonumber \\
&-2b_ik_p\beta_3-2\overline d_{2i} \overline d_{4i})\omega^6+(\overline d^2_{4i}+2\overline d_{2i}-b_i^2k_v^2 \nonumber \\
&-2\overline d_{3i}\overline d_{5i})\omega^8+(\overline d^2_{5i}-2\overline d_{4i})\omega^{10}+\omega^{12}\geq 0,\;\forall\; \omega\in\mathbb{R}.
\end{align}
Through calculation, we know
\begin{align*}
&x_{di}^2(\omega)+y_{di}^2(\omega)-x_{ni}^2(\omega)-y_{ni}^2(\omega)\nonumber\\
=&(2b_ik_p\beta_3 \overline n_{2i}+{\overline d^2_{1i}}-2b_ik_p\beta_3\overline d_{2i}-\overline n^2_{1i})\omega^2+(2\overline n_{1i}\overline n_{3i} \nonumber \\
\hspace{-4mm}&+2b_ik_p\beta_3\overline d_{4i}+\overline d^2_{2i}-\overline n^2_{2i}-2b_i^2k_pk_v\beta_3-2\overline d_{1i}\overline d_{3i})\omega^4\nonumber \\
\hspace{-4mm}&+(2\overline n_{2i}b_ik_v+2\overline d_{1i}\overline d_{5i}+\overline d^2_{3i}-\overline n^2_{3i}-2b_ik_p\beta_3 \nonumber \\
\hspace{-4mm}&-2\overline d_{2i} \overline d_{4i})\omega^6+(\overline d^2_{4i}+2\overline d_{2i}-b_i^2k_v^2-2\overline d_{3i}\overline d_{5i})\omega^8 \nonumber \\
\hspace{-4mm}&+(\overline d^2_{5i}-2\overline d_{4i})\omega^{10}+\omega^{12}
\end{align*}
This together with (\ref{equ1e}) leads to
\begin{align}\label{equxe}
x_{di}^2(\omega)+y_{di}^2(\omega)-x_{ni}^2(\omega)-y_{ni}^2(\omega)\geq 0,\;\forall\; \omega\in\mathbb{R}.
\end{align}
By (\ref{equxe}), we know
\begin{align}\label{equx1e}
\frac{x_{ni}^2(\omega)+y_{ni}^2(\omega)}{x_{di}^2(\omega)+y_{di}^2(\omega)}\leq1,\;\forall\; \omega\in\mathbb{R}.
\end{align}
Through calculation, we obtain
\begin{align*}
\left|\frac{x_{ni}(\omega)+y_{ni}(\omega)j}{x_{di}(\omega)+y_{di}(\omega)j}\right|=
\frac{\sqrt{x_{ni}^2(\omega)+y_{ni}^2(\omega)}}{\sqrt{x_{di}^2(\omega)+y_{di}^2(\omega)}}.
\end{align*}
This together with (\ref{equx1e}) leads to
\begin{align}\label{equx2e}
\left|\frac{x_{ni}(\omega)+y_{ni}(\omega)j}{x_{di}(\omega)+y_{di}(\omega)j}\right| \leq 1,\;\forall\; \omega\in\mathbb{R}.
\end{align}
From (\ref{G_ee}) and (\ref{equx2e}), we know $|G_{ei}(j\omega)|\leq1$ for any $\omega\in \mathbb{R}$. That is $\|G_{ei}(s)\|_{\infty}\leq1$.\qed

\end{appendices}

\bibliographystyle{model5-names}        % Include this if you use bibtex
\bibliography{LLGD2020Automatica_arXiv-Final}           % and a bib file to produce the
% bibliography (preferred). The
% correct style is generated by
% Elsevier at the time of printing.

%\begin{thebibliography}{99}     % Otherwise use the
% thebibliography environment.
% Insert the full references here.
% See a recent issue of Automatica
% for the style.
%  \bibitem[Heritage, 1992]{Heritage:92}
%     (1992) {\it The American Heritage.
%     Dictionary of the American Language.}
%     Houghton Mifflin Company.
%  \bibitem[Able, 1956]{Abl:56}
%     B.~C.~Able (1956). Nucleic acid content of macroscope.
%     {\it Nature 2}, 7--9.
%  \bibitem[Able {\em et al.}, 1954]{AbTaRu:54}
%     B.~C. Able, R.~A. Tagg, and M.~Rush (1954).
%     Enzyme-catalyzed cellular transanimations.
%     In A.~F.~Round, editor,
%     {\it Advances in Enzymology Vol. 2} (125--247).
%     New York, Academic Press.
%  \bibitem[R.~Keohane, 1958]{Keo:58}
%     R.~Keohane (1958).
%     {\it Power and Interdependence:
%     World Politics in Transition.}
%     Boston, Little, Brown \& Co.
%  \bibitem[Powers, 1985]{Pow:85}
%     T.~Powers (1985).
%     Is there a way out?
%     {\it Harpers, June 1985}, 35--47.

%\end{thebibliography}
\end{document}